\documentclass[prd,preprint,superscriptaddress,nofootinbib,longbibliography,tightenlines]{revtex4-1}
\usepackage{graphicx}
\usepackage{amsmath}
\usepackage{amsfonts}
\usepackage{mathtools}
\usepackage{color}
\usepackage{amssymb}
\usepackage{dcolumn}
\usepackage{bm}
\usepackage{braket}
\usepackage{microtype} 
\usepackage[linktoc=all]{hyperref}
\usepackage{cleveref}

\numberwithin{equation}{section}

\newcommand{\ud}{\mathrm{d}}

\newcommand{\mn}{{\mu\nu}}
\newcommand{\pp}{\pi_\phi}

\newcommand{\bel}[1] {\begin{equation}\label{#1}}
\newcommand{\beal}[1] {\begin{eqnarray}\label{#1}}

\newcommand{\be}{\begin{equation}}
\newcommand{\bea}{\begin{eqnarray}}
\newcommand{\ee}{\end{equation}}
\newcommand{\eea}{\end{eqnarray}}

\makeatletter
\DeclareRobustCommand{\rcite}[1]{%
	\rcite@aux#1,\@nil{#1}%
}
\def\rcite@aux#1,#2\@nil#3{%
	\if\relax#2\relax
	Ref.~\cite{#3}%
	\else
	Refs.~\cite{#3}%
	\fi
}
\makeatother

\begin{document}
\title{The WKB approximation and tunneling in theories with non-canonical kinetic terms}
	
	\author{Mariana Carrillo Gonz\'alez} 
	\email{cmariana@sas.upenn.edu}
	\affiliation{Center for Particle Cosmology, Department of Physics and Astronomy,
		University of Pennsylvania, Philadelphia, Pennsylvania 19104, USA}
	\author{Ali Masoumi}
	\email{ali.masoumi@tufts.edu}
	\affiliation{Institute of Cosmology, Department of Physics and Astronomy,
		Tufts University, Medford, MA 02155, USA}
	\author{Adam R. Solomon}
	\email{adamsol@physics.upenn.edu}
	\affiliation{Center for Particle Cosmology, Department of Physics and Astronomy,
		University of Pennsylvania, Philadelphia, Pennsylvania 19104, USA}
	\author{Mark Trodden}
	\email{trodden@physics.upenn.edu}
	\affiliation{Center for Particle Cosmology, Department of Physics and Astronomy,
		University of Pennsylvania, Philadelphia, Pennsylvania 19104, USA}
	
	\date{\today}
	
\begin{abstract}
Tunneling is a fascinating aspect of quantum mechanics that renders the local minima of a potential meta-stable, with important consequences for particle physics, for the early hot stage of the universe, and more speculatively, for the behavior of the putative multiverse. While this phenomenon has been studied extensively for systems which have canonical kinetic terms, many theories of fundamental physics contain fields with non-canonical kinetic structures. It is therefore desirable to have a detailed framework for calculating tunneling rates and initial states after tunneling for these theories. In this work we present such a rigorous formulation and illustrate its use by applying it to a number of examples.
\end{abstract}
	
\maketitle

\tableofcontents

\section{Introduction}
The process of quantum tunneling allows transitions out of local minima of an energy functional to vacua of lower (or higher) energies. This occurs through a first-order phase transition mediated by the nucleation of bubbles of the new vacuum inside the old. This process can have important consequences, not only for particle physics, but also for cosmology since, starting from its hot initial state, the universe may have gone through several of these phase transitions before it settled into its current vacuum.  It is quite possible that the initial state for inflation may have been set by the state after quantum tunneling. It is even possible that our own vacuum may be susceptible to such transitions. Indeed, the Higgs potential with the currently accepted values of top quark and Higgs masses is metastable and, in the absence of new physics, can decay, albeit after a rather long time (see, for example, \rcite{Sher:1988mj}).

Decay rates are (almost) always calculated in a semiclassical regime using the WKB approximation. The generalization of the WKB approximation to cases with more than one degree of freedom was first presented in \rcite{BBW, Banks:1974ij}. This was extended to field theories in several important works \cite{Kobzarev:1974cp, Coleman:1977py, Callan:1977pt} and later to cases which include gravitational back-reaction on tunneling \cite{Coleman:1980aw, Hawking:1981fz}. Some analytic approximations for  tunneling rates in  thin-wall regime were devised in \rcite{Coleman:1977py, Parke:1982pm}. 

The possibility of the existence of the string landscape, and the attendant possibility of many phase transitions in such a complex potential has attracted further interest in vacuum decay processes. However, despite progress in understanding vacuum tunneling, our only analytic insight, through the thin-wall approximation, is solely applicable to cases where the tunneling action is large, and as such is only relevant to a very specific class of real-world processes. Accounting for gravity and spacetime curvature brings about a new set of problems. There are many conceptual complications in the presence of gravity, such as the measure problem (see \rcite{Freivogel:2011eg} for a review) or the interpretation of Hawking-Moss channels of tunneling \cite{Weinberg:2006pc}. A further computational issue is that we do not know whether the tunneling rate in the presence of gravity is dominated by solutions which are $O(4)$ symmetric in Euclidean space, despite some effort in this direction \cite{Masoumi:2012yy, Garriga:2004nm}. Furthermore, the string landscape and most other putative landscapes usually have a large number of fields. The process of tunneling here is plagued with many computational difficulties, although these were recently circumvented in an efficient numerical package \cite{Masoumi:2016wot}. 

Our goal in this paper is to provide a careful analysis of another important issue in a number of models relevant to cosmology, that of the problem of tunneling in theories with non-canonical kinetic terms (for a review of such models, see \rcite{Clifton:2011jh,Joyce:2014kja}). These theories appear in many cases in modern cosmological models, and, as we shall see, the decay rate can be highly non-intuitive (see also \rcite{Brown:2007ce}).

This paper is organized as follows. In \cref{sec:wkb} we study a general formalism for the WKB approximation for arbitrary Hamiltonians in quantum mechanics, and in \cref{sec:DecayRates} we calculate decay rates. We generalize these results to quantum field theory in \cref{sec:Field}, and provide several applications of our results in \cref{sec:Applications} before concluding in \cref{sec:Conclusion}. 

\section{WKB for arbitrary Hamiltonians} \label{sec:wkb}
In a system described by a Hamiltonian $H(\mathbf{q},\mathbf{p})$, we can find the classical motion by solving the Hamilton-Jacobi equation,
\begin{equation}
H(\mathbf{q},\nabla S)+\frac{\partial S}{\partial t}=0 \ ,
\end{equation}
where $\mathbf{q}=(q_1,\cdots,q_n)$ are the coordinates, $\mathbf{p}=(p_1,\cdots,p_n)$ are the canonical momenta, and $S$ is the Hamilton principal function given by 
\begin{equation}
S(\mathbf{q},\boldsymbol{\alpha};t)=\int^{\mathbf{q}}\mathbf{p}(\mathbf{q'},\boldsymbol{\alpha})\cdot\ud\mathbf{q'}-\int H\ud t \ , 
\label{se}
\end{equation}
satisfying $\nabla S=\mathbf{p}$. The corresponding quantum system is described by the Hamiltonian operator $\hat{H}$, related to the classical one by
\begin{equation}
\hat{H} = \frac{1}{(2\pi\hbar)^{2n}} \int \ud \mathbf{p} \, \ud \mathbf{q} \, \ud \mathbf{u} \, \ud \mathbf{v}\: F(\mathbf{u}\cdot\mathbf{v}/ \hbar) H(\mathbf{q}, \mathbf{p}, t) e^{(i/\hbar)[(\mathbf{q}-\hat{\mathbf{Q}})\cdot\mathbf{u} + (\mathbf{p} - \hat{\mathbf{P}})\cdot\mathbf{v}]} \ ,
\end{equation}
where $\hat{\mathbf{Q}}$,$\hat{\mathbf{P}}$ are the coordinate and momentum operators respectively, and $F(\mathbf{u}\cdot\mathbf{v}/ \hbar)$ is the transformation function~\cite{Cohen:1966} that defines the operator ordering and must be real in order to ensure that $\hat{H}$ is Hermitian.\footnote{If the Hamiltonian is not Hermitian, an extra exponential term appears in the WKB wave function \cite{Miz81}.} This Hamiltonian appears in the Schr\"{o}dinger equation that describes the quantum system,
\begin{equation}
i\hbar\frac{\partial \psi(\mathbf{q}, t)}{\partial t}=\hat{H}\left(\mathbf{q}, -i \hbar \frac{\,\partial}{\partial\mathbf{q}};t\right) \psi(\mathbf{q}, t) \ .
\end{equation}
The semi-classical solution, often referred to as the WKB approximation, for this equation up to $\mathcal{O}(\hbar)$ is given by\footnote{As long as $F(0)=1$ and $F'(0)=0$, which is satisfied for the most common transformation functions \cite{Miz77}.} \cite{Miz77,Brown72}
\begin{equation}
\psi(\mathbf{q}, t)=N \sqrt{\det{\left(\frac{\partial^2 S}{\partial \mathbf{q} \partial \boldsymbol{\alpha}}\right)}}e^{i/\hbar\; S(\mathbf{q} \ ,\boldsymbol{\alpha};t)},
\end{equation}
where $N$ is a normalization constant,  $\alpha_i$ with $i=1,\cdots,n$ are integration constants\footnote{In the Hamilton-Jacobi formalism these are the new momenta; the fact that they are constant in time follows from the requirement that the transformed Hamiltonian be identically zero.} that are determined by the initial conditions. We may fix the first constant as $\alpha_1=E$, while the remaining $\alpha_i$'s are chosen depending on the system at hand. For example, if $H=H(x)+H(y)+H(z)$, we can pick $\alpha_1=E_\text{tot}=H$, $\alpha_2=E_\text{x}=H(x)$, and $\alpha_3=E_\text{y}=H(y)$, whereas if we have spherical symmetry, then some of the $\alpha_i$'s will correspond to angular momenta. We can see that the time-independent wave function is approximated in the semi-classical limit as
\begin{equation}
\psi(\mathbf{q})=N \sqrt{\det{\left(\frac{\partial^2 S}{\partial \mathbf{q} \partial \boldsymbol{\alpha}}\right)}}e^{i/\hbar\; \int^\mathbf{q}\mathbf{p}(\mathbf{q'},\boldsymbol{\alpha})\cdot\ud\mathbf{q'}} \ . \label{psi}
\end{equation}
To $\mathcal{O}(\hbar^0)$, we may neglect the pre-factor in \cref{psi}, keeping only the leading-order exponential behavior. This can be understood more easily by recalling that WKB is a semi-classical approximation in $\hbar$; that is, $\psi_\text{WKB}=e^{i(\sigma_0+\hbar \sigma_1)/\hbar}$. The order $\hbar^0$ factor is 
\begin{equation}
\sigma_0\equiv i \int^\mathbf{q}\mathbf{p}(\mathbf{q'},\boldsymbol{\alpha})\cdot\ud\mathbf{q'} \ ,
\end{equation}
while the order $\hbar$ contribution, $\sigma_1$, is logarithmic and gives rise to the aforementioned pre-factor.
\begin{figure}[!t]
	\includegraphics[scale=0.9]{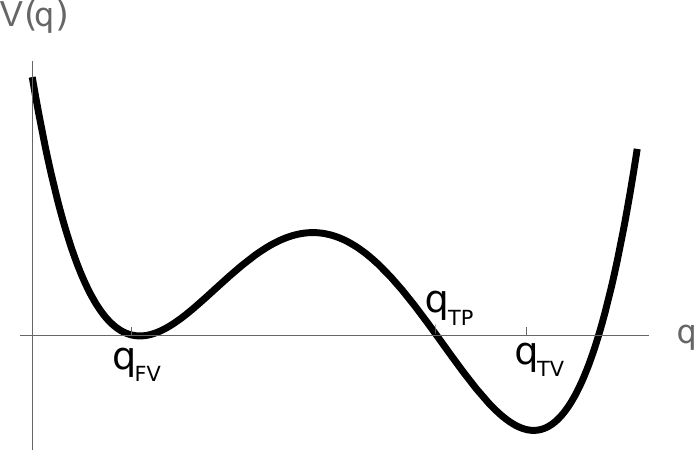}
	\caption{An example of a potential where tunneling can happen from the false vacuum $q_\mathrm{FV}$ to the true vacuum $q_\mathrm{TV}$.}
	\label{fig:pot}
\end{figure}
The WKB approximation is widely used to solve tunneling problems. The one-dimensional case is straightforward, since there is only one tunneling path to follow. The multi-dimensional case becomes more complicated due to the different paths through which tunneling is possible. Banks, Bender, and Wu \cite{BBW} solved this problem by considering the most probable escape paths (MPEPs), which are expected to dominate the amplitude. From \cref{psi} we can see that the largest contribution to the amplitude comes from paths which minimize the WKB exponent, i.e., the MPEPs are the paths that satisfy
\begin{equation}
\delta\int_{\mathbf{q}_\text{FV}}^{\mathbf{q}_\text{TP}} \,\mathbf{p}\cdot\ud\mathbf{q}=0 \ , \label{dp}
\end{equation}
where $\mathbf{q}_\text{FV}$ and $\mathbf{q}_\text{TP}$ are the locations of the false vacuum and the turning point, defined by $V(\mathbf{q}_\mathrm{FV}) = V(\mathbf{q}_\mathrm{TP})=E$; a typical setup (compressed to one dimension) is illustrated in \cref{fig:pot}. In the classically forbidden region, through which tunneling occurs, $\mathbf{p}$ is imaginary and thus the wave function decays exponentially. \Cref{dp} can be written in a more suggestive form by using the definition of a MPEP. In order to do so, consider~\cite{Bitar:1978vx,Copeland:2007qf} a curve $\mathbf{Q}(\lambda)$ parametrized by $\lambda$ and notice that, in the classically-forbidden region, we have
\begin{equation}
\nabla \sigma_0 \cdot \nabla \sigma_0 = |\mathbf{p}|^2 \ , \label{eq:gradB2}
\end{equation}
where the gradient is taken with respect to $\mathbf{q}$. We can expand the gradient in terms of the tangent vector to the curve $\mathbf{Q}$, $\mathbf{v_\parallel}=\partial \mathbf{Q}/\partial \lambda$, and the vectors orthogonal to $\mathbf{Q}$,  $\mathbf{v}^i_\perp$, as
\begin{equation}
\left.\nabla\right|_{\mathbf{q}=\mathbf{Q}} = \frac{\mathbf{v_\parallel}}{|\mathbf{v_\parallel}|^2}\left.\left(\mathbf{v_\parallel}\cdot\nabla\right)\right|_{\mathbf{q}=\mathbf{Q}} + \displaystyle\sum_i \frac{\mathbf{v}^i_\perp}{|\mathbf{v}^i_\perp|^2}\left.\left(\mathbf{v}^i_\perp\cdot\nabla\right)\right|_{\mathbf{q}=\mathbf{Q}} \ .
\end{equation}
This decomposition is useful here because MPEPs are defined as the paths that satisfy
\begin{equation}
\left.\mathbf{v}^i_\perp\cdot\nabla \sigma_0\right|_{\mathbf{q}=\mathbf{Q}}=0\qquad\forall i \ .
\end{equation}
To take advantage of this, let us reparametrize the curve as $\mathbf{Q}(\lambda(s))$, with $s$ the proper distance along $\mathbf Q$,
\begin{equation}
\ud s=|\ud\mathbf{Q}|=\sqrt{\frac{\ud \mathbf{Q}}{\ud \lambda}\cdot\frac{\ud \mathbf{Q}}{\ud \lambda}}\ud \lambda=|\mathbf{v_\parallel}|\ud \lambda \ ,
\end{equation}
so that we have
\begin{equation}
\left.\nabla \sigma_0\right|_{\mathbf{q}=\mathbf{Q}} = \frac{\mathbf{v_\parallel}}{|\mathbf{v_\parallel}|^2}\left.\left(\mathbf{v_\parallel}\cdot\nabla \sigma\right)\right|_{\mathbf{q}=\mathbf{Q}} = \frac{\mathbf{v_\parallel}}{|\mathbf{v_\parallel}|}\frac{\ud \sigma}{\ud s} \ .
\end{equation}
Using this in \cref{eq:gradB2}, we can finally rewrite \cref{dp} as
\begin{equation}
\delta\int_{s(\mathbf{q}_\text{FV})}^{s(\mathbf{q}_\text{TP})} |\mathbf{p}(\mathbf{Q}(s),E)| \ud s=0 \ , \label{varp}
\end{equation}
where $|\mathbf{p}|$ is found by solving $H(\mathbf{q},\mathbf{p})=E$. The variation in \cref{varp} keeps the starting point fixed but not the end point,\footnote{This is because in the multidimensional case there is generally not a single point $\mathbf{q}_\mathrm{TP}$ but rather a surface of points satisfying the condition $V(\mathbf{q}_\mathrm{TP})=E$.} with energy conserved along the path. The fact that the endpoint is not fixed gives rise to the boundary condition
\begin{equation}
\left.\frac{\ud \mathbf{q}}{\ud \lambda}\right|_{\mathbf{q}=\mathbf{q}_\text{TP}}=0 \ .
\end{equation}
Now let us choose the parameter $\lambda$ such that
\begin{equation}
\frac{\ud s}{\ud \lambda}=\left|\frac{\partial H}{\partial \mathbf{p}}\right| \ ,
\end{equation}
in which case \cref{varp} translates to
\begin{equation}
\frac{\ud}{\ud \lambda}\left(\frac{|\mathbf{p}|}{\left|\frac{\partial H}{\partial \mathbf{p}}\right|}\frac{\ud \mathbf{Q}}{\ud \lambda}\right)-\left|\frac{\partial H}{\partial \mathbf{p}}\right|\nabla|\mathbf{p}|=0 \ . \label{mpe}
\end{equation}

In the following, we will assume that there is a well-defined Legendre transformation that allows us to switch between the Hamiltonian and Lagrangian formulations. A careful analysis, taking into account that we are in the classically forbidden region, shows that \cref{mpe} can be written as
\begin{equation}
\frac{\ud}{\ud \lambda}\left(\frac{\partial L_\text{E}}{\partial \frac{\ud \mathbf{Q}}{\ud \lambda}}\right)-\frac{\partial L_\text{E}}{\partial \mathbf{Q}}=0 \ ,
\end{equation}
where we have used Hamilton's equations and $L_\mathrm{E}$ is the Euclidean Lagrangian. This shows that the MPEP can be found by solving the \emph{Euclidean} equations of motion. Note that, since analytic continuation can lead to multi-valued functions, the MPEP $\mathbf{Q}(\lambda)$ could be multi-valued. 

The fact that the MPEP can be found by solving the Euclidean equations of motion has previously been shown for canonical kinetic terms and here we have extended the proof for generic kinetic terms of the form $T(\mathbf{q},\dot{\mathbf{q}})$. \emph{That this result applies for generic kinetic terms $T(\mathbf{q},\dot{\mathbf{q}})$ is one of the main results of this paper.} Later, we will show that this result also holds for scalar fields with second-order equations of motion.

\section{Computing the decay rate}\label{sec:DecayRates}

Once we have an approximation for the wave function, we can use it to calculate the decay rate in a potential with two non-degenerate minima as in \cref{fig:pot}. The decay rate of a system is defined as
\begin{equation}
\Gamma=-\frac{1}{P_\text{FV}}\frac{\ud}{\ud t}P_\text{FV} \ ,
\end{equation}
where $P_\text{FV}$ is the probability of being in the false vacuum. As discussed in \rcite{Masoumi:2015psa,Andreassen:2016cvx}, this definition is only meaningful for times $t_\text{slosh}\ll t\ll t_\text{non-lin}$, where $t_\text{slosh}=\omega_\text{FV}^{-1}$ with $\omega_\text{FV}$ the frequency of oscillation in the false vacuum and $t_\text{non-lin}$ the scale at which non-linearities become important. During $t<t_\text{slosh}$, high energy modes in the initial wave function will decay, and it is not until these modes decay that we truly observe the decay rate of the false vacuum. We may write the decay rate as
\begin{equation}
\Gamma=\frac{1}{m}\frac{\int|\psi_E(\mathbf{q}_\text{TP})|^2\mathbf{p}_\text{TP}\cdot \ud \mathbf{q}_\text{TP}}{\int_\text{FV}\ud \mathbf{q} |\psi_E(\mathbf{q})|^2} \ , \label{dr}
\end{equation}
where TP is the turning point, with the integration over all possible turning points, FV stands for the false vacuum, and $\psi_E$ is an energy eigenstate. Using the WKB approximation up to $\mathcal{O}(\hbar)$, this translates to
\begin{equation}
\Gamma=\left.\frac{\left. \det\left(\frac{\partial ^2S}{\partial \boldsymbol{\alpha}\partial\mathbf{q}}\right)\right|_{\mathbf{q}=\mathbf{q}_\text{TP}}|\mathbf{p}_\text{TP}|}{m\int_0^{\mathbf{q}_\text{FV}} \det{\left(\frac{\partial ^2S}{\partial \boldsymbol{\alpha}\partial\mathbf{q}}\right)}\ud \mathbf{q}}e^{-B}\right|_{\mathbf{q}=\mathbf{Q}} \ , \quad B\equiv\frac{2i}{\hbar}\; \int_{\mathbf{q}_\text{FV}}^{\mathbf{q}_\text{TP}}\mathbf{p}\cdot\ud\mathbf{q} \ , \label{gamma}
\end{equation}
where $\mathbf{Q}$ is the MPEP and $B$ is the WKB exponent. For a canonical kinetic term in one dimension we have 
\begin{equation}
\det{\left(\frac{\partial ^2S}{\partial {\alpha}\partial{q}}\right)}=\frac{2m}{|p|} \ ,
\end{equation}
which leads to the well-known result
\begin{equation}
\Gamma=\left.\frac{|p_\text{FV}|}{m|q_\text{FV}|}e^{-B}\right|_{q=Q} \ .
\end{equation}
In the case of a canonical kinetic term, the pre-factor has a clear physical interpretation: writing it as $v_\text{FV}/|\mathbf{q}_\text{FV}|$, it can be understood as the rate at which the wave function hits the barrier. However, for the case of non-canonical kinetic terms, it is not simple to find a similar interpretation, and the rest of this paper will be concerned solely with the exponent $B$.
\begin{figure}[!t]
	\includegraphics[scale=0.6]{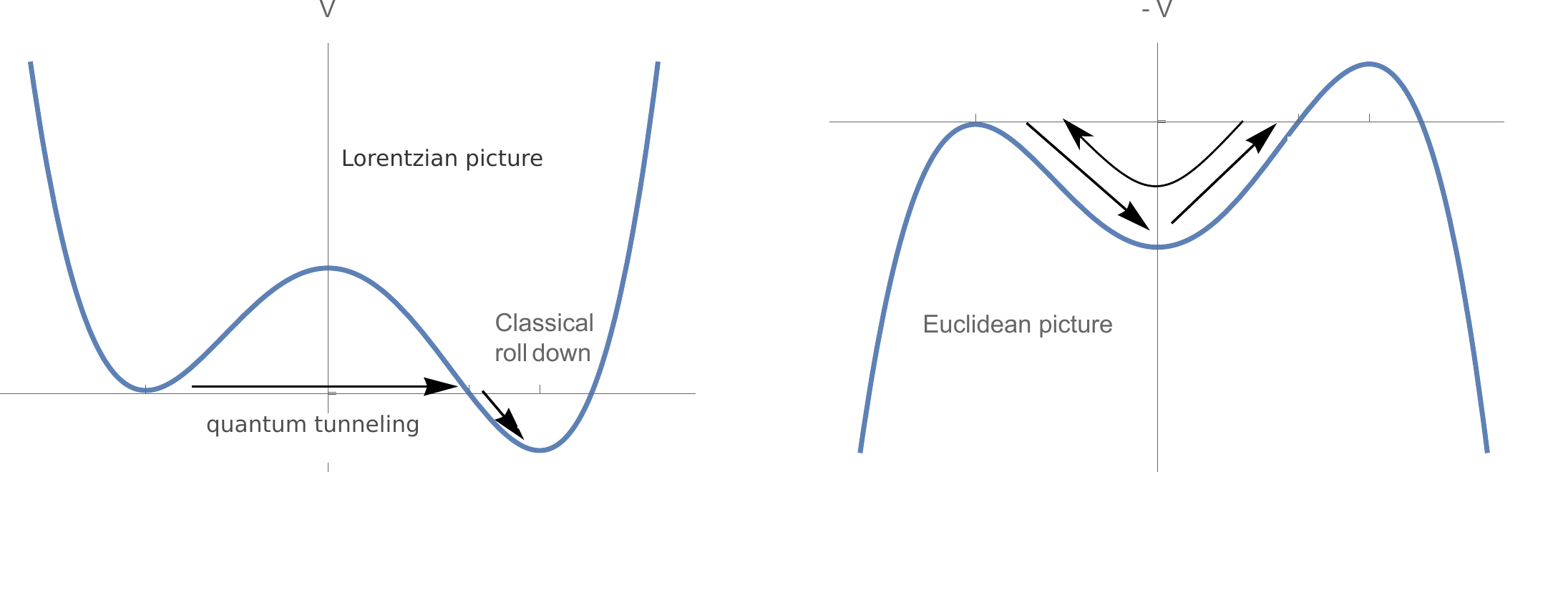}
	\caption{Lorentzian and Euclidean pictures of the false vacuum tunneling; in the Euclidean picture the potential is inverted.}
	\label{fig:instbou}
\end{figure}

We now review the calculation of the WKB exponent to leading order for the tunneling of the false vacuum, illustrated in \cref{fig:instbou}.
The tunneling between an unstable vacuum and a lower energy (local or global) vacuum, is commonly called the \textit{bounce}, which is just a specific kind of instanton\footnote{In general, an instanton is a configuration with a finite, non-zero action that solves the classical equations of motion.}. The WKB exponent is calculated by solving the Euclidean equations of motion, i.e., with the potentials inverted. In the bounce potential, the particle rolls up to the turning point and then falls back down to the false vacuum (this is, of course, the origin of the term ``bounce" for this process).

To relate the WKB exponent with the Euclidean action $S_\mathrm{E} = iS$, we begin by using \cref{se}, which tells us that 
\begin{equation}
iS(\mathbf{q})=i\int^\mathbf{q}\mathbf{p}\cdot\ud\mathbf{q'}+iS(\mathbf{q_\text{FV}}) \ ,
\end{equation}
where we have used the fact that the kinetic energy at $\mathbf{q}_\text{FV}$ vanishes, so that we can set
\begin{equation}
\int H \ud t=\int V(\mathbf{q_\text{FV}}) \ud t=-S(\mathbf{q}_\text{FV}) \ ,
\end{equation}
where $H=E$ is conserved. Given this and being careful with the integration limits, we can write the exponent $ B$ for the tunneling of the false vacuum as 
\begin{equation}
B_\text{bounce}=S_\mathrm{E}(\mathbf{q})-S_\mathrm{E}(\mathbf{q_\text{FV}}) \ .\label{bou}
\end{equation}
It is important to note that this relation only holds at stationary points of $ B$ (and $S$), i.e., when the equations of motion are satisfied. One should realize that the path in Euclidean space goes from $\mathbf{q_\text{FV}}$ at $\tau=-\infty$ to $\mathbf{\mathbf{q}_\text{TP}}$ at a finite $\tau$ (which can generally be taken to be $\tau=0$) and back to $\mathbf{q_\text{FV}}$ at $\tau=\infty$; this path gives the correct factor in \cref{bou}.
Given this, the unstable vacuum decay rate to $\mathcal{O}(\hbar^0)$ is written as
\begin{equation}
\Gamma=\left.e^{-B}\right|_{\mathbf{q}=\mathbf{Q}}=\left.e^{-\frac{1}{\hbar}
\left(S_\mathrm{E}(\mathbf{q})-S_\mathrm{E}(\mathbf{q_\text{FV}})\right)}\right|_{\mathbf{q}=\mathbf{Q}} \ ,
\end{equation}
which is a well-known result. 

\section{WKB in a general scalar quantum field theory}\label{sec:Field}
In this section we generalize the results obtained in \cref{sec:wkb,sec:DecayRates} for multi-dimensional quantum mechanics to quantum field theory with a scalar field, again closely following \rcite{Bitar:1978vx, Copeland:2007qf}. Crucially, we will allow for a general enough kinetic structure for our formalism to cover all Lorentz-invariant scalar-field theories with equations of motion that are second order, and therefore avoid the Ostrogradski ghost instability. As discussed in \cref{app:gals}, the Lagrangians for these theories take the form (up to boundary terms) $L = L(\phi,\dot\phi,\nabla\phi,\nabla^2\phi)$, where $\nabla^2\phi=\partial_i\partial_j\phi$ is a matrix (rather than the scalar Laplacian). Defining the canonical momentum as usual, $\Pi=\ud L/\ud\dot\phi$, we can therefore write the Hamiltonian in the form
\begin{equation}
H=\int\ud^3q\left[T(\phi,\Pi,\nabla\phi,\nabla^2\phi) + G(\nabla\phi,\nabla^2\phi)+V(\phi)\right] \ . \label{eq:QFTham}
\end{equation}

Consider a wave functional $\psi[\phi]$, a functional of $\phi(q)$ whose squared norm is the probability density for a configuration $\phi(q)$. This will obey the generalized Schr\"{o}dinger equation
\begin{equation}
\left[\int\ud^3q\,T\left(\phi,-i\hbar\frac{\delta}{\delta\phi(q)},\nabla\phi,\nabla^2\phi \right)+U[\phi]\right]\psi[\phi]=E\psi[\phi] \ , \label{gensch}
\end{equation}
where the functional $U[\phi]$ is the potential energy that determines the possibility of tunneling, defined by
\begin{equation}
U[\phi]=\int\ud^3q\left(G(\nabla\phi,\nabla^2\phi )+V(\phi)\right)\ .
\end{equation}
The classically-forbidden region is given by $E<U[\phi]$. The configuration space is the space of real-valued functions on $\mathbb{R}^3$ (or the relevant space depending on the problem at hand) satisfying the appropriate boundary conditions.

We proceed to make a semi-classical approximation as in the quantum mechanics case; to do so, we expand the wave function as
\begin{equation}
\psi[\phi]=e^{\frac{i}{\hbar}\sigma[\phi]}=e^{\frac{i}{\hbar}\left(\sigma_0[\phi]+\hbar\,\sigma_1[\phi]+\cdots\right)} \ .
\end{equation}
In the following, we solve for the wave function to $\mathcal{O}(\hbar^0)$. Substituting the semi-classical expansion in \cref{gensch} gives, at leading order,
\begin{equation}
\int\ud^3q\,T\left(\phi,\frac{\delta \sigma_0[\phi]}{\delta\phi(\mathbf{q})},\nabla\phi,\nabla^2\phi \right)+U[\phi]=E \ .
\end{equation}
Expressing the canonical momentum as a function of $\phi$ and its gradients, by making use of the conservation of energy equation, we find that the leading-order contribution is
\begin{equation}
\sigma_0[\phi]=\int^\phi\ud\phi'\;\Pi(\phi',\nabla\phi',\nabla^2\phi' ) \ .
\end{equation}

The next step is to find the MPEP, i.e., the curve in the space of real-valued functions (or field configurations) that minimizes $\sigma_0[\phi]$. We will call this curve $\Phi(\lambda,\mathbf{q})$, parametrized by $\lambda$, denote the vector parallel to this curve by $\mathbf{v}_\parallel(\lambda,\mathbf{q})=\partial\Phi/\partial \lambda$, and label the continuous set of perpendicular vectors $\mathbf{v}_\perp(\lambda,\mathbf{q}_1;\mathbf{q}_2)$. In this case, the condition defining the MPEP is
\begin{equation}
\int\ud^3q\,\mathbf{v}_\perp(\lambda,\mathbf{q}_1;\mathbf{q}_2)\left.\frac{\delta \sigma_0[\phi]}{\delta\phi(\mathbf{q}_1)}\right|_{\phi=\Phi}=0 \quad\forall\mathbf{q}_2 \ .
\end{equation}
We reparametrize the curve as $\Phi(\lambda(s),\mathbf{q})$, with $s$ the proper distance along the curve, given by
\begin{equation}
\ud s=|\ud\Phi|=\sqrt{\int\ud^3q\,\left(\frac{\ud \Phi}{\ud \lambda}\right)^2}\ud \lambda \ .
\end{equation}
Using this parametrization, we find that the MPEP satisfies
\begin{equation}
\delta\int\Pi(\Phi,\nabla\Phi,\nabla^2\Phi ) \ud s=0 \ , \label{varf}
\end{equation}
and since
\begin{equation}
\frac{\ud s}{\ud \lambda}=\left|\frac{\partial H}{\partial \Pi}\right| \ ,
\end{equation}
we then find that \cref{varf} translates into
\begin{equation}
\frac{\ud}{\ud \lambda}\left(\frac{\Pi}{\left|\frac{\partial H}{\partial \Pi}\right|}\frac{\ud \Phi}{\ud \lambda}\right)+\nabla\left(\left|\frac{\partial H}{\partial \Pi}\right|\frac{\partial\Pi}{\partial\nabla\Phi}\right)-\nabla^2\left(\left|\frac{\partial H}{\partial \Pi}\right|\frac{\partial\Pi}{\partial\nabla^2\Phi}\right)-\left|\frac{\partial H}{\partial \Pi}\right|\frac{\partial\Pi}{\partial\Phi}=0 \ . 
\end{equation}
This is again equivalent to finding the Euclidean equations of motions,
\begin{equation}
\frac{\ud}{\ud \lambda}\left(\frac{\partial L_\text{E}}{\partial \frac{\ud \Phi}{\ud \lambda}}\right)+\nabla\left(\frac{\partial L_\text{E}}{\partial \nabla\Phi}\right)-\nabla^2\left(\frac{\partial L_\text{E}}{\partial\nabla^2\Phi}\right)-\frac{\partial L_\text{E}}{\partial \Phi}=0 \ ,
\end{equation}
that is, the MPEP is a stationary solution of the Euclidean action. The generalization of this calculation to include higher-order gradients in $T$ and $G$ is straightforward, although we remind the reader that for the most general scalar field theories with second-order equations of motion, these terms only depend on spatial gradients up to $\nabla^2\phi$. We conclude that even in the presence of non-canonical kinetic terms, the dominant contribution to the tunneling rate comes from paths which extremize the Euclidean action.

\section{Application: Decay rates in general scalar-field theories}\label{sec:Applications}
To this point we have established a rigorous formalism for computing the decay rates for tunneling processes in scalar field theories with kinetic terms of the form $T(\phi,\dot{\phi},\nabla\phi,\nabla^2\phi,\cdots)$. In this section we explicitly compute decay rates for general theories of a single scalar field with second-order equations of motion, known broadly as galileons.\footnote{These theories are introduced in \cref{app:gals}; in particular, the galileons Lagrangians are given by \cref{eq:gal-lag}. We emphasize that these Lagrangians completely cover theories of a single scalar field on a flat background with second-order equations of motion.} We find a simple and familiar expression for the decay rate, and discuss how decay of the false vacuum could occur considerably more quickly than in theories with just a canonical kinetic term.

Consider a scalar field $\phi$ defined on flat space and endowed with a potential $V(\phi)$ with two minima, one at slightly higher potential than the other, as shown in \cref{fig:pot}. A state localized in the false vacuum, denoted by $V_\text{FV}$, can decay to the true vacuum at $V_\text{TV}$. We will denote the value of $\phi$ at these minima by $\phi_+$ and $\phi_-$, respectively. We have shown above that, regardless of the choice of kinetic term, the decay rate per unit volume for this process is given by
\begin{equation}
\frac\Gamma V \sim e^{-B} \ ,
\end{equation}
where $B\equiv\Delta S_\mathrm{E}$ is the difference between the Euclidean action for two different solutions: a ``bounce," in which the scalar field rolls from the true vacuum to the false vacuum,\footnote{The field rolls from true vacuum to false because motion in Euclidean time can be thought of as motion in the inverted potential.} and a solution in which the field lives at the false vacuum for all time. The analysis in this section largely follows the classic work of Coleman \cite{Coleman:1977py}.

Before diving into general cases, with all their attendant abstraction, let us start by considering a particularly simple example of a non-canonical kinetic term: $P(X)$ theories, with an action of the form\footnote{In fact, we can consider a function $P(\phi,X)$ without affecting our results; however, for clarity we will start off by cleanly separating the kinetic and potential terms. The more general case is discussed later in this section.}
\begin{equation}
S = \int\ud^4x\left[P(X)+V(\phi)\right] \ ,
\end{equation}
where $X\equiv-(\partial\phi)^2$ and we assume (without loss of generality) that $P(0)=0$. The conditions for the bounce are consistent with an $O(4)$-symmetric solution for $\phi$ \cite{Coleman:1977py}, so the bounce solution is generally taken to have this symmetry.\footnote{For a canonical kinetic term it can be proven that $e^{-B}$ is extremized for an $O(4)$-symmetric solution \cite{Coleman:1977th}, though no such proof currently exists for non-canonical terms.} For a solution with this symmetry, the Euclidean action is given by
\begin{equation}
S_\mathrm{E} = 2\pi^2\int\rho^3\left(P + V\right)\ud\rho \ ,
\end{equation}
with $X=\dot\phi^2$, where $\rho$ is the Euclidean $O(4)$ radial coordinate.

Following \rcite{Brown:2007ce} (in which tunneling was studied in a particular $P(X)$ theory) we will make a slightly non-standard definition of $L$ as the Lagrangian with the spherical measure factor divided out,
\begin{equation}
S_\mathrm{E} \equiv 2\pi^2\int \rho^3L\ud\rho \ ,
\end{equation}
and define a similarly non-standard canonical momentum as
\begin{equation}
\pp \equiv \frac{\partial L}{\partial \dot\phi} \ .
\end{equation}
Using as our Lagrangian $L=P+V$, the canonical momentum is
\begin{equation}
\pi_\phi = \frac{\partial L}{\partial \dot\phi} = 2P_X\dot\phi = 2P_X\sqrt X \ ,
\end{equation}
so that the Hamiltonian defined with respect to this $L$ is
\begin{equation}
H = \pp\dot\phi - L = 2P_XX - P - V \ .
\end{equation}
This Hamiltonian is not conserved, since the spherical measure induces a friction term in the equation of motion. The ``true" conserved Hamiltonian is $\rho^3H$, whose associated canonical momentum is $\partial (\rho^3L)/\partial\dot\phi = \rho^3\pp$. Hamilton's equations then imply
\begin{align}
\dot\pi_\phi &= -\frac{\partial H}{\partial \phi} - \frac3\rho\pp \ .
\end{align}

Now let us consider the bounce and false-vacuum solutions for $\phi$ in the \emph{thin-wall approximation} in which
\begin{equation}
\epsilon\equiv V_\text{FV} - V_\text{TV}
\end{equation}
is small. In this approximation the thickness of the wall is very small compared to the radius of the wall, $\bar\rho$, which we can define as the point at which $\phi(\bar\rho) = \frac12(\phi_++\phi_-)$. Moreover, in this limit our nonstandard $H$ is approximately conserved: since the field should be stationary in the two vacua, the difference in $H$ from one side of the wall to the other should just be proportional to the difference in the potentials, and therefore to $\epsilon$. Accordingly we can write $H + \mathcal{O}(\epsilon) = E$ for a conserved $E$, implying
\begin{equation}
2P_XX - P = E + V + \mathcal{O}(\epsilon) \ . \label{eq:PXcons}
\end{equation}
We may obtain the energy $E$ by evaluating this for $\rho>\bar\rho$, where both the bounce solution and the always-false-vacuum solution are in the false vacuum, $\phi=\phi_+$. Since $\dot\phi$ has to vanish at this point in both solutions, and $P(0)=0$ by construction, the left-hand side vanishes, so we have $E=-V_\text{FV}$. We can simplify this further by defining a new function, $V_0(\phi)$, as a deformation of the potential which vanishes, along with its first derivative, at the two vacua, i.e.,
\begin{equation}
V_0(\phi) \equiv V(\phi) - V_\text{FV} + \mathcal{O}(\epsilon),\qquad V_0(\phi_\pm) = V_0'(\phi_\pm)=0 \ .
\end{equation}
Up to $\mathcal{O}(\epsilon)$ we may simply replace the right-hand side of \cref{eq:PXcons} with $V_0(\phi)$,
\begin{equation}
2P_XX - P = V_0 + \mathcal{O}(\epsilon) \ . \label{eq:PXcons2}
\end{equation}

To calculate the bounce factor $B$,
\begin{equation}
B = S_\mathrm{E}(\phi) - S_\mathrm{E}(\phi_+) \ ,
\end{equation}
with $\phi$ the bounce solution, we split the computation up into three different regions: in the true vacuum, in  the false vacuum, and on the wall, i.e.,
\begin{equation}
B = B_\mathrm{FV} + B_\mathrm{wall} + B_\mathrm{TV} \ .
\end{equation}
Equivalently, this can be thought of as splitting the integrals into pieces from 0 to $\bar \rho$ (the true vacuum), near $\bar\rho$ (the wall), and from $\bar\rho$ to $\infty$ (the false vacuum).

In the false vacuum we simply have $B_\mathrm{FV} = \left.\left[S_\mathrm{E}(\phi_+) - S_\mathrm{E}(\phi_+)\right]\right|_{\bar\rho}^\infty = 0$, where in each $S_E$ we are only integrating from $\rho=\bar\rho$ to $\rho=\infty$. In the true vacuum,
\begin{align}
B_\mathrm{TV} &= 2\pi^2\int_0^{\bar\rho}\rho^3(V_\text{FV}-V_\text{TV})\ud\rho = -\frac{\pi^2\bar\rho^4}{2}\epsilon \ .
\end{align}
Finally, on the wall we have $\rho\approx\bar\rho$, so that in the thin-wall approximation, $\int_\mathrm{wall}\rho^3\ud\rho f(\phi,\dot\phi) = \bar\rho^3\int_{\phi_-}^{\phi_+}\ud\phi f/\dot\phi$, for some generic function $f(\phi,\dot\phi)$. We can then calculate the portion of $B$ on the wall as
\begin{align}
B_\mathrm{wall} &= 2\pi^2\bar\rho^3\int_{\phi_-}^{\phi_+}\frac{P+V}{\dot\phi}\ud\phi - 2\pi^2\bar\rho^3\int_{\phi_-}^{\phi_+}\frac{V_\text{FV}}{\dot\phi} \ud\phi = 2\pi^2\bar\rho^3S_1 \ ,
\end{align}
to leading order in $\epsilon$, where
\begin{equation}
S_1 \equiv \int_{\phi_-}^{\phi^+} \pp\ud\phi \label{eq:PXS1}
\end{equation}
is the tension of the bubble wall. Putting all these together we find the well-known result \cite{Coleman:1977py},
\begin{equation}
B = 2\pi^2\bar\rho^3S_1 - \frac{\pi^2\bar\rho^4}{2}\epsilon \ .
\end{equation}
We can determine $\bar\rho$ by demanding that it extremize $B$; i.e. that $\partial B/\partial\bar\rho = 0$, yielding
\begin{equation}
 \bar\rho = \frac{3S_1}{\epsilon} \ .
\end{equation}
This gives us the usual result,
\begin{equation}
B = \frac{27\pi^2S_1^4}{2\epsilon^3} \ . \label{eq:PXB}
\end{equation}

Our main result for decay rates in $P(X)$ theories, summarized in \cref{eq:PXS1,eq:PXB}, reduces to the classic result when we choose a canonical kinetic term \cite{Coleman:1977py}, and also includes the results of \rcite{Brown:2007ce}, which studied the case of a Dirac-Born-Infeld (DBI) kinetic term, which is a $P(X)$ theory with $P(X) \sim f^{-1}(\sqrt{1+fX}-1)$. As pointed out in \rcite{Brown:2007ce}, despite the cosmetic similarities between the tunneling rates for various kinetic terms, even small changes can have a tremendous impact. The choice of kinetic term modifies the wall tension $S_1$ (cf. \cref{eq:PXS1}). This shows up in the tunneling rate as $e^{-(\cdot\cdot\cdot) S_1^4}$, so minor alterations to the kinetic structure of a theory can affect its tunneling rate by several orders of magnitude.

We can see this explicitly by solving for $\pp$ using the conservation equation for the Hamiltonian,
\begin{equation}
H = V_0 + \mathcal{O}(\epsilon) \ ,
\end{equation}
in order to determine $S_1$ in terms of $P(X)$ and $V_0(\phi)$. For example, taking a canonical kinetic term, $P(X) = X/2$, we have $\pp = \dot\phi = \sqrt{2V_0}$, leading to
\begin{equation}
S_1 = \int_{\phi_-}^{\phi^+} \sqrt{2V_0} \ud\phi \ ,
\end{equation}
which appears in the standard result for the tunneling rate \cite{Coleman:1977py}. The analogous result for a general $P(X)$ is obtained by solving \cref{eq:PXcons2} for $\pp=2P_X\sqrt X$. This can lead to important changes in $\pp$ and therefore, through $S_1$, in the decay rate $\Gamma$. We emphasize that by phrasing our result in terms of the non-standard canonical momentum $\pp$, we can write the decay rate for $P(X)$ theories in a simple form that incorporates both the classic result for a canonical kinetic term \cite{Coleman:1977py} as well as more recent extensions \cite{Brown:2007ce}.

Now let us add one layer of abstraction by considering a general Lagrangian depending on $\phi$ and $\dot\phi$; in practice this amounts to a $P(X)$ theory with $\phi$ dependence, but it will prove a useful arena for building a more abstract calculation of the decay rate which we can then apply to the general second-order scalars.

Energy conservation gives 
\begin{equation}
H = \pp\dot\phi - L = E + \mathcal{O}(\epsilon) \ ,
\end{equation}
and by evaluating this expression at $\phi=\phi_+$ we find $E=-L(\phi_+,0)$. Since $\dot\phi=0$ at this point, $E$ is a constant and can be thought of as analogous to $-V_\mathrm{FV}$. Calculating $B$ in three parts as above, we find $B_\mathrm{FV}=0$,
\begin{equation}
B_\mathrm{TV} = 2\pi^2\int_0^{\bar\rho}\rho^3\left[L(\phi_-,0) - L(\phi_+,0)\right]\ud\rho = -\frac{\pi^2\bar\rho^4}{2}\epsilon \ ,
\end{equation}
where we have defined $\epsilon \equiv L(\phi_+,0) - L(\phi_-,0)$, and
\begin{align}
B_\mathrm{wall} &= 2\pi^2\bar\rho^3\int_{\phi_-}^{\phi+}\pp\ud\phi \equiv 2\pi^2\bar\rho^3S_1 \ ,
\end{align}
with the rest of the calculation of $\Gamma/V$ following as above. We conclude that for a general Lagrangian depending on $\phi$ and $\dot\phi$, the tunneling rate is given by a simple generalization of the classic result,
\begin{equation}
\frac\Gamma V \sim e^{-B} \ ,
\end{equation}
where
\begin{align}
B &= \frac{27\pi^2S_1^4}{2\epsilon^3} \ , \\
S_1 &= \int_{\phi_-}^{\phi^+} \pp\ud\phi \ .
\end{align}

Finally, let us extend our calculation of the decay rate to the full set of scalar field theories with second-order equations of motion, the well-known galileons and their generalizations. While these Lagrangians can depend on second derivatives of $\phi$ in specific, antisymmetric combinations (cf. \cref{eq:gal-lag}), integrations by parts can eliminate the dependence of $L$ on $\ddot\phi$ at the expense of introducing explicit $\rho$ dependence, as shown explicitly in \cref{app:gals}. This is a consequence of the galileon structure, which ensures that the equations of motion are second-order, and would not remain true for Lagrangians with general functions of $\partial^2\phi$.

We can therefore consider the full slate of healthy theories of a single scalar field by generalizing the above analysis to $L=L(\phi,\dot\phi,\rho)$. We will assume that $L$ loses its $\rho$ dependence when $\dot\phi=0$, i.e.,
\begin{equation}
\frac{\partial L(\phi,0,\rho)}{\partial\rho}=0 \ ,
\end{equation}
as this holds for the galileons and rather simplifies the analysis.\footnote{When $\dot\phi=0$, we will write quantities with two arguments rather than three, e.g., writing $L(\phi,0,\rho)$ as $L(\phi,0)$, reflecting the fact that such objects do not in fact depend on $\rho$.} Note that this implies that, away from the wall, $L$ is constant.

Most of the features of the above calculation proceed practically unchanged by the additional $\rho$ dependence in $L$, 
with the final result taking the form
\begin{align}
B &= 2\pi^2\bar\rho^3S_1 - \frac{\pi^2\bar\rho^4}{2}\epsilon \ , \label{eq:b}\\
S_1 &= \int_{\phi_-}^{\phi^+} \pp(\phi,\dot\phi,\rho)\ud\phi \ .
\end{align}
On the face of it, the entire structure of the decay rate up to this point is unaffected by the $\rho$ dependence. However, the crucial difference is that $S_1$ now depends on $\bar\rho$, so that when we calculate $\bar\rho$ by minimizing $B$, as above, we will find that the structure of $S_1$ can play an additional role, since $\partial B/\partial\bar\rho=0$ now yields
\begin{align}
3S_1 - \epsilon\bar\rho + \bar\rho\frac{\partial S_1}{\partial\bar\rho} &= 0 \ . 
\label{eq:rhobarsol}
\end{align}

As a concrete example, consider the cubic galileon with a canonical kinetic term,
\begin{equation}
S_\mathrm{E} = \int\ud^4x\left[\frac12(\partial\phi)^2 + \frac{1}{\Lambda^3}(\partial\phi)^2\Box\phi + V(\phi)\right] \ ,
\end{equation}
corresponding to
\begin{equation}
L = \frac12\dot\phi^2 + \frac{2}{\Lambda^3}\frac{\dot\phi^3}{\rho} + V \ .
\end{equation}
The canonical momentum is
\begin{equation}
\pp = \dot\phi + \frac{6}{\Lambda^3}\frac{\dot\phi^2}{\rho} \ ,
\end{equation}
so that the surface tension of the bubble wall is
\begin{equation}
S_1(\bar\rho) = \int_{\phi_-}^{\phi^+} \left(\dot\phi + \frac{6}{\Lambda^3}\frac{\dot\phi^2}{\bar\rho}\right)\ud\phi\equiv S_1^\mathrm{can}+\frac{1}{\bar{\rho}}S_1^\mathrm{gal} \ ,
\end{equation}
where $S_1^\mathrm{can}$ and $S_1^\mathrm{gal}$ are defined so as not to depend on $\bar\rho$. Plugging this into \cref{eq:b}, we can minimize $B$ to find $\bar\rho$ as usual,
\begin{equation}
\bar\rho = \frac{3S_1^\mathrm{can}}{2\epsilon}\left[1 + \sqrt{1+\frac89\lambda}\right]\ ,
\end{equation}
where we have defined
\begin{equation}
\lambda \equiv \frac{S_1^\mathrm{gal}\epsilon}{(S_1^\mathrm{can})^2}\ .
\end{equation}
Substituting this back into \cref{eq:b} we find
\begin{equation}
B = \frac{27\pi^2(S_1^\mathrm{can})^4}{\epsilon^3}\Delta^3\left(1+\frac23\frac\lambda\Delta\right)\ , \label{eq:gal-B}
\end{equation}
with
\begin{equation}
\Delta\equiv\frac12\left(1+\sqrt{1+\frac89\lambda}\right)\ .
\end{equation}

In these expressions for $\bar\rho$ and $B$ we have not yet taken a thin-wall limit, and it is not hard to see why: the correct limit to take depends on whether the canonical term or the galileon dominates $S_1$, i.e., whether
\begin{equation}
\frac{S_1^\mathrm{gal}\epsilon}{(S_1^\mathrm{can})^2} \gg 1 \ ,\qquad\text{or}\qquad \frac{S_1^\mathrm{gal}\epsilon}{(S_1^\mathrm{can})^2} \ll 1\ .
\end{equation}
This depends on the free parameters of the theory: $\epsilon$, which controls the difference between the potentials of the two vacua; $\Delta\phi\equiv\phi_+-\phi_-$, the difference between the field values at the two vacua; and $\Lambda$, which controls the size of the galileon term. Given these parameters, we can estimate the dominant contribution to $S_1$ as follows. Let us approximate the field profile as $\phi\simeq \frac{\Delta\phi}{2}\tanh{(\frac{\Delta\phi}{2}(\rho-\bar\rho))}$; while this simple ansatz will not exactly solve the equations of motion (although it does in the absence of the galileon and in the limit $\epsilon\to0$ \cite{Coleman:1977py}), in the thin-wall limit we expect qualitatively similar behavior, so our choice will be sufficient to relate $\lambda$ to the theory parameters. Evaluating this field profile on $S_1^\mathrm{can}$ and $S_1^\mathrm{gal}$, we find
\begin{equation}
\lambda \equiv \frac{S_1^\mathrm{gal}}{(S_1^\mathrm{can})^2}\epsilon=6\frac{\epsilon }{\Delta \phi \, \Lambda ^3} \ .
\end{equation}
We see that the canonical kinetic term dominates the decay rate if $\frac{\epsilon}{\Delta\phi}\ll\Lambda^3$, and the galileon dominates the rate if $\frac{\epsilon}{\Delta\phi}\gg\Lambda^3$. Note that $\frac{\epsilon}{\Delta\phi}=\frac{\Delta V}{\Delta \phi}$ is the overall slope of the potential between the two vacua.

\begin{figure}[!t]
	\includegraphics[scale=0.54]{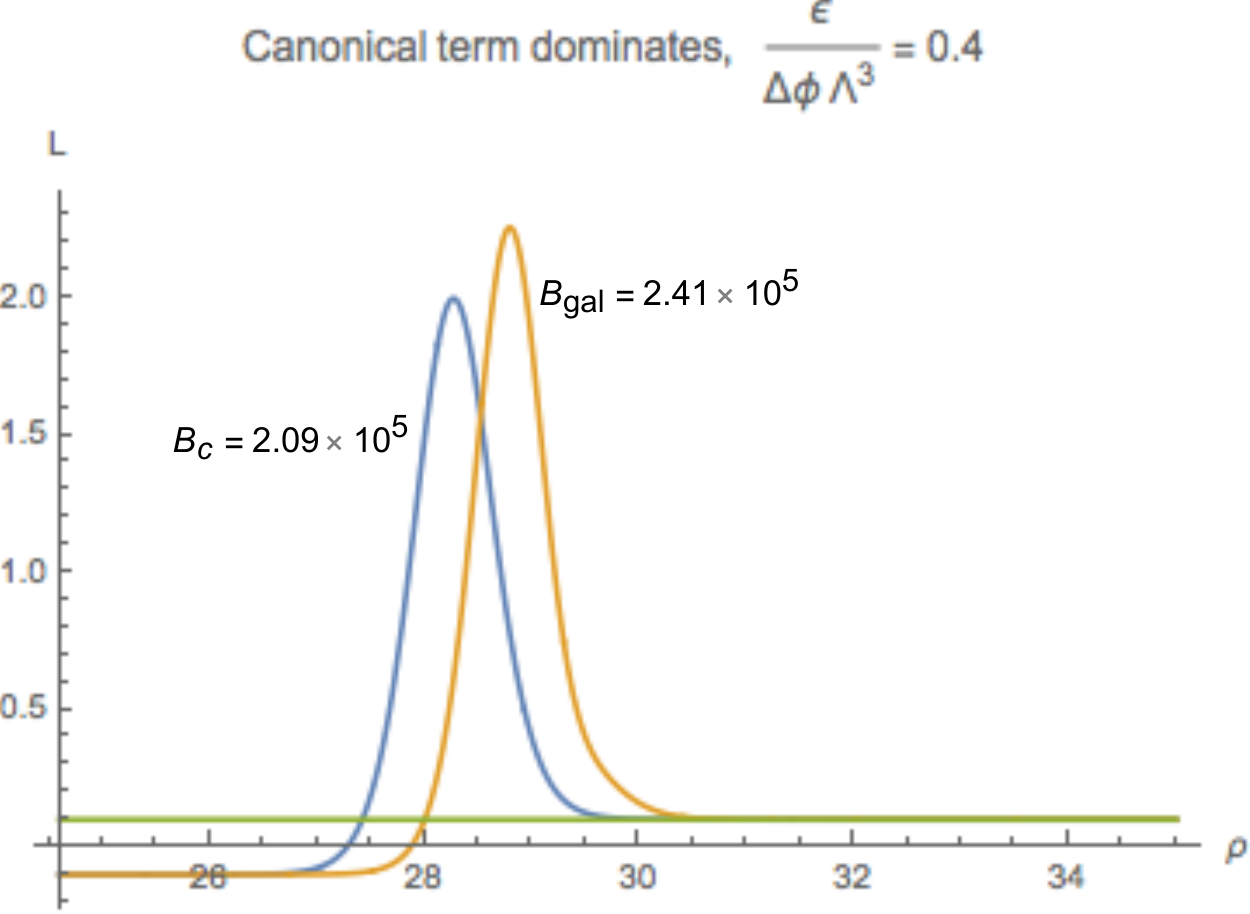}
	\includegraphics[scale=0.54]{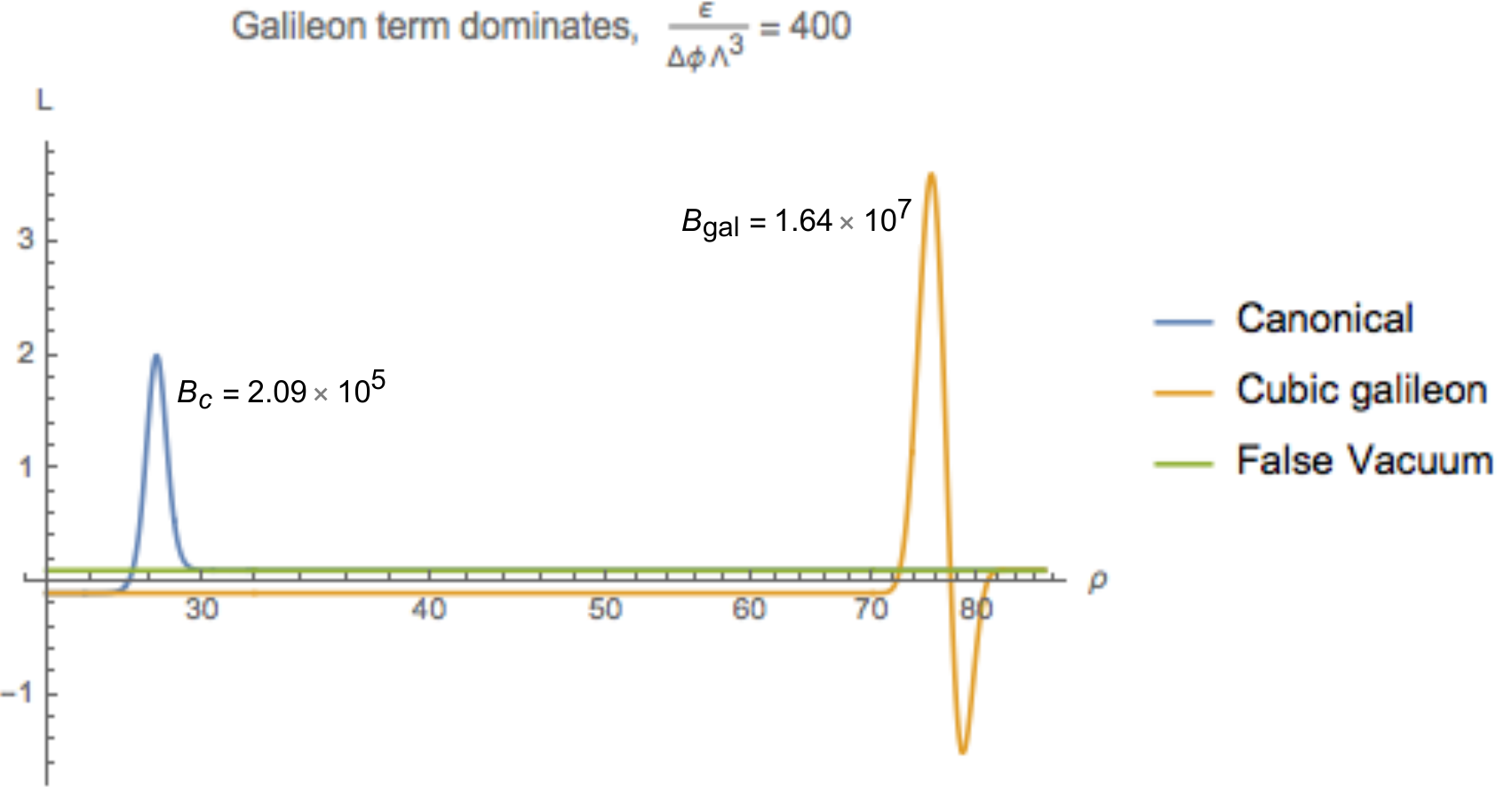}
	\caption{In this figure, we observe the behavior of $L(\phi_\text{bounce})$ for two different limits. On the left side we see the case where the canonical term dominates and on the right side the case where the galileon term dominates. $B_\text{can}$ and $B_\text{gal}$ are the WKB exponents for a canonical scalar field and for a canonical + cubic galileon scalar field respectively. It is clear that, when the galileon term dominates we see a drastic change in the decay rate.}
	\label{fig:galbou}
\end{figure}

We are now in a position to take the thin-wall limit and evaluate the decay rate in the presence of a cubic galileon. In the limit where the canonical kinetic term dominates we have the usual decay rate,
\begin{equation}
B_\mathrm{can} = \frac{27\pi^2(S_1^\text{can})^4}{2\epsilon^3} \ ,
\end{equation}
while when the galileon dominates, we find
\begin{equation}
B_\mathrm{gal}=\frac{2 \pi ^2 (S_1^{\mathrm{gal}})^2}{\epsilon } \ .
\end{equation}
In \cref{fig:galbou}, we can observe the change of the WKB exponent in both limits, when the canonical term dominates and when the galileon term dominates. We see that the change in the decay rate will be drastic when the galileon term dominates. We conclude that the galileon can lower the decay rate, potentially by a rather large amount, compared to a canonical scalar.

In fact, we can apply this reasoning to the full range of galileons (and therefore of healthy scalar theories). It is not too difficult to show that a general galileon Lagrangian, allowing for all the galileon terms with any functions of $\phi$ and $X$ in front, leads to a Euclidean action of the form
\begin{equation}
S_\mathrm{E} = 2\pi^2\int\rho^3L\ud\rho \ ,
\end{equation}
with
\begin{equation}
L = \displaystyle\sum_{n=0}^3\frac{f_n(\phi,\dot\phi)}{\rho^n} \ .
\end{equation}
Note that $f_0$ receives contributions from $P(X)$ terms and the cubic galileon, $f_1$ from the cubic and quartic terms, $f_2$ from the quartic and quintic terms, and $f_3$ from the quintic term. The bubble tension is
\begin{equation}
S_1 = \displaystyle\sum_{n=0}^3\frac{g_n(\dot\phi)}{\bar\rho^n} \ ,
\end{equation}
where we have defined
\begin{equation}
g_n = \int_{\phi_-}^{\phi_+}\frac{\partial f_n}{\partial\dot\phi}\ud\phi \ .
\end{equation}
Solving for $\bar\rho$ by minimizing $B$ we obtain
\begin{equation}
\epsilon\bar\rho + \displaystyle\sum_{n=0}^3\frac{(n-3)g_n}{\bar\rho^n} = 0 \ .
\end{equation}
Note that the $n=3$ piece does not contribute, so (after multiplying by $\bar\rho^2$) this is a cubic equation for $\bar\rho$,
\begin{equation}
\epsilon\bar\rho^3 - 3g_0\bar\rho^2 - 2g_1\bar\rho - g_2 = 0 \ .
\end{equation}
We have already addressed above the special case where $g_2=0$ and this equation is quadratic, i.e., when only the cubic galileon and a $P(X)$ term are present. If this equation is cubic, we can perform a similar analysis; solving for $B$ we find
\begin{equation}
B = \frac{27\pi^2g_0^4}{2\epsilon^3}\left[1+\frac{4}{3}\frac{g_1}{g_0^2}\epsilon+\frac{4  \left(3 g_0 g_2+2 g_1^2\right)}{27 g_0^4}\epsilon ^2+\cdots\right] \ ,\label{eq:b-full}
\end{equation}
and while neglecting higher-order terms in $\epsilon$ is tempting, the same lesson we learned above holds: if the $g_n$ terms, with $n\geq1$, are larger than $g_0$, one should keep a different set of terms in \cref{eq:b-full}. In this case, we have three expansion parameters given by 
\begin{equation}
 \frac{g_n}{g_0^{n+1}} \epsilon ^n,\quad n=1,2,3 \ ,
\end{equation}
for which an analysis similar to the cubic galileon one can be performed, given a specific action.

\section{Discussion}
\label{sec:Conclusion}
Scalar field theories with non-canonical kinetic structures play an important role in building phenomenologically interesting models of both the early and late universe. Some classes of such theories arise naturally in supergravity and string theory, and others arise as limits of massive gravity and brane-world constructions. In each case, it is interesting to wonder whether the nonperturbative physics of these theories might provide a novel way to constrain and test them, and whether they can yield results significantly different from canonical fields.

In this paper we have examined tunneling in general scalar field theories, allowing for the existence of non-canonical kinetic structures, while demanding the the resulting equations of motion be second order, and hence ghost-free. We have shown how to construct the general tunneling formalism for such theories and applied it to several well-known examples, in the thin-wall limit. While the formal structure of the expressions for the decay rates are the same for both these theories and for canonical ones, the resulting tunneling rates can be dramatically altered by the presence of non-canonical terms, giving rise to significant differences in the decay rates.

%
%
%
%

\begin{acknowledgments}
We are thankful to Garrett Goon, Kurt Hinterbichler, Austin Joyce, Matteo Vicino, Alex Vilenkin and Erick Weinberg for useful comments and discussion. Calculations have made use of xAct.\footnote{http://www.xact.es/} The work of M.C. and M.T was supported in part by NASA ATP grant NNX11AI95G. The work of A.M. was supported by NSF grant: PHY-1518742. The work of A.R.S. was supported by funds provided to the Center for Particle Cosmology by the University of Pennsylvania. M.T. was also supported in part by US Department of Energy (HEP) Award DE-SC0013528.

\end{acknowledgments}

\appendix
\section{Galileons}
\label{app:gals}

In this appendix we briefly discuss the \emph{galileons} \cite{Nicolis:2008in} and their generalizations, which are the most general scalar field theories with second-order equations of motion \cite{Horndeski:1974wa,Deffayet:2011gz}, and show how they fit into the formalisms discussed in \cref{sec:Field,sec:Applications}.

Constructing a scalar field theory on flat space and insisting that the equations of motion be second order in derivatives,\footnote{This requirement is necessary to avoid the Ostrogradsky instability \cite{Ostrogradsky:1850fid,Woodard:2006nt}. This may be loosened somewhat when multiple fields are present \cite{deRham:2016wji}, as in the so-called ``beyond-Horndeski" theories \cite{Gleyzes:2014dya,Zumalacarregui:2013pma} and their generalizations \cite{Crisostomi:2016czh,Achour:2016rkg}, but for a single scalar field this loophole is not available.} we are led to the Lagrangian \cite{Horndeski:1974wa,Deffayet:2011gz}
\begin{equation}
\mathcal{L} = \displaystyle\sum_{n=2}^DG_n(\phi,X)\mathcal{L}_n
\end{equation}
in $D$ spacetime dimensions, where $G_n(\phi,X)$ are arbitrary functions of $\phi$ and $X=-(\partial\phi)^2$. In $n=4$ the individual galileon Lagrangians are
\begin{align}
\mathcal{L}_2 &= (\partial\phi)^2, \nonumber \\
\mathcal{L}_3 &= (\partial\phi)^2\Box\phi, \nonumber \\
\mathcal{L}_4 &= (\partial\phi)^2\left[(\Box\phi)^2 - \phi_\mn^2\right], \nonumber \\
\mathcal{L}_5 &= (\partial\phi)^2\left[(\Box\phi)^3 - 3\Box\phi \phi_\mn^2 + 2\phi_\mn^3\right], \label{eq:gal-lag}
\end{align}
where we have defined $\phi_\mu\equiv\partial_\mu\partial_\nu\phi$. We will frequently refer to $\mathcal{L}_3$, $\mathcal{L}_4$, and $\mathcal{L}_5$ as the cubic, quartic, and quintic galileons, respectively.

First we will justify the form \eqref{eq:QFTham} of the Hamiltonian we considered, in which dependence on higher spatial gradients but not on higher time derivatives is permitted. \textit{A priori} it is not obvious that the Hamiltonians for the galileons (above $\mathcal{L}_2$) fall into that class, since the Lagrangians themselves, in their covariant form, contain second derivatives of $\phi$. However, the fact that the resulting equations of motion are second order ensures that we are able to eliminate higher time derivatives up to boundary terms.

As an illustration, consider the cubic galileon,\footnote{In the above notation, this corresponds to $G_3$ constant.} with the action
\begin{equation}
S = \int\ud^4x\left(-\frac12(\partial\phi)^2 + \frac{1}{\Lambda^3}(\partial\phi)^2\Box\phi - V(\phi)\right) \ ,
\end{equation}
where $\Lambda$ is a constant with units of mass. Performing a $3+1$ spacetime decomposition we have
\begin{align}
S &= \int\ud t\ud^3x \left[\frac12\left(\dot\phi^2 - (\nabla\phi)^2\right) + \frac{1}{\Lambda^3}\left(-\dot\phi^2 + (\nabla\phi)^2\right)\left(-\ddot\phi + \nabla^2\phi\right) - V\right] \nonumber \\
&\equiv \int\ud t \ud^3 x L \ .
\end{align}
We may then eliminate the $\ddot\phi$ dependence by integrating by parts. Consider the term
\begin{equation}
\left(-\dot\phi^2 + (\nabla\phi)^2\right)\left(-\ddot\phi + \nabla^2\phi\right) = \dot\phi^2\ddot\phi - \ddot\phi(\nabla\phi)^2 - \dot\phi^2\nabla^2\phi + (\nabla\phi)^2\nabla^2\phi \ .
\end{equation}
The first piece is a total derivative in time, $\dot\phi^2\ddot\phi = \frac13(\dot\phi^3)\dot{}$. The next term can be eliminated by a pair of total derivatives,
\begin{equation}
\frac{\ud}{\ud t}\left[\dot\phi(\nabla\phi)^2\right] - \partial_i\left[\dot\phi^2\partial^i\phi\right] = \ddot\phi(\nabla\phi)^2 - \dot\phi^2\nabla^2\phi \ ,
\end{equation}
leaving us with
\begin{equation}
L = \frac12\left(\dot\phi^2 - (\nabla\phi)^2\right) + \frac{1}{\Lambda^3}\left(-2\dot\phi^2 + (\nabla\phi)^2\right)\nabla^2\phi - V \ ,
\end{equation}
up to boundary terms. We can therefore obtain the canonical momentum,
\begin{equation}
\Pi = \frac{\ud L}{\ud \dot\phi} = \dot\phi\left(1-\frac{4}{\Lambda^3}\nabla^2\phi\right) \ ,
\end{equation}
and solve for $H(\phi,\Pi,\partial_i\phi,\partial_i\partial_j\phi)$. The cubic galileon therefore fits into the form used in \cref{sec:Field}.

This property has also been shown to apply to the quartic and quintic galileons~\cite{Sivanesan:2011kw}. To round out the list of second-order field theories, we only need to generalize this to include $\phi$ and $X$-dependence in the coefficients $G_n$. For simplicity, let us look at the cubic galileon with some general $\phi$- and $X$-dependent coefficient,
\begin{equation}
\mathcal{L} = e^{\alpha(\phi,X)}\Box\phi \ .
\end{equation}
We will find it convenient to explicitly consider how $\alpha$ separately depends on $\dot\phi$ and $\partial_i\phi$,
\begin{equation}
\alpha(\phi,X)\to\alpha(\phi,\rho,\partial_i\phi) \ ,
\end{equation}
where for further convenience we have defined $\rho = \log (\dot\phi / \Lambda^2)$, with $\Lambda$ a constant with dimensions of mass. Using
\begin{equation}
X = -\dot\phi^2 + \partial_i\phi\partial^i\phi \ ,
\end{equation}
we see that, of course, derivatives of $\alpha$ with respect to $\rho$ and $\partial_i\phi$ are related to each other,
\begin{align}
\alpha_\rho &= -2\alpha_X\dot\phi^2 \ , \\
\alpha_i &= 2\alpha_X\partial_i\phi \ , \label{eq:alph-i}
\end{align}
where we have defined
\begin{equation}
\alpha_\rho \equiv \frac{\partial\alpha}{\partial\rho},\quad\alpha_i \equiv \frac{\partial\alpha}{\partial\partial^i\phi},\quad\alpha_X \equiv \frac{\partial\alpha}{\partial X}.
\end{equation}
We now write the Lagrangian explicitly in terms of time and space derivatives,
\begin{equation}
\mathcal{L} = -e^\alpha\ddot\phi + e^\alpha\nabla^2\phi \ .
\end{equation}
The second term is already of the form we want: it depends only on $\phi$, $\dot\phi$, and spatial derivatives of $\phi$ (but \emph{not} of $\dot\phi$). We now work on the first term. Integrating by parts on the time derivative, and rearranging, we have
\begin{equation}
-e^\alpha\ddot\phi \sim \frac{e^\alpha}{1+\alpha_\rho}\left(\alpha_\phi\dot\phi^2 + \alpha_i\dot\phi\partial^i\dot\phi\right) \ ,
\end{equation}
where $\sim$ denotes equivalence up to boundary terms.This explains our choice to use $\alpha$ and $\rho$ rather than $G_3$ and $\dot\phi$. The first term in this expression is of the form we want, but we need to remove the spatial derivative from $\dot\phi$ in the second term. Taking this term separately,  integrating by parts on the spatial derivative, and using (cf. \cref{eq:alph-i}), that
\begin{equation}
\alpha_{\rho i} = \frac{\partial \alpha_\rho}{\partial\partial^i\phi} = 2\alpha_{\rho X}\partial_i\phi = \frac{\alpha_{\rho X}}{\alpha_X}\partial_i\phi \ ,
\end{equation}
we obtain
\begin{align}
\frac{e^\alpha}{1+\alpha_\rho}\alpha_i\dot\phi\partial^i\dot\phi &\sim -\frac{e^\alpha}{2+3\alpha_\rho+\alpha_\rho^2-\alpha_{\rho\rho}+\frac{\alpha_{\rho X}}{\alpha_X}(1+\alpha_\rho)} \nonumber \\
&\hphantom{=} \times \dot\phi^2\left[ \left(\alpha_{\phi i}+\alpha_\phi\alpha_i - \frac{\alpha_{\phi\rho}\alpha_i}{1+\alpha_\rho}\right)\partial^i\phi + \left(\alpha_{ij} + \alpha_i\alpha_j - \frac{\alpha_{\rho j}\alpha_i}{1+\alpha_\rho}\right)\partial^i\partial^j\phi\right] \ .
\end{align}
Similar proofs apply to the quartic and quintic galileons multiplied by general functions.

A similar (and more straightforward) calculation justifies the formalism used in \cref{sec:Applications} to compute Euclidean bounce solutions; in particular, the Euclidean action for $O(4)$-symmetric solutions can be written in the form
\begin{equation}
S_\mathrm{E} = 2\pi^2\int\rho^3L(\phi,\dot\phi,\rho)\ud\rho \ ,
\end{equation}
where the $\rho$ dependence in $L(\phi,\dot\phi,\rho)$ comes only from the cubic, quartic, and quintic galileons after integrating by parts, and
\begin{equation}
\frac{\partial L(\phi,0,\rho)}{\partial\rho}=0 \ .
\end{equation}

\bibliographystyle{apsrev4-1}
\bibliography{bibliography}

\begin{thebibliography}{37}%
\makeatletter
\providecommand \@ifxundefined [1]{%
 \@ifx{#1\undefined}
}%
\providecommand \@ifnum [1]{%
 \ifnum #1\expandafter \@firstoftwo
 \else \expandafter \@secondoftwo
 \fi
}%
\providecommand \@ifx [1]{%
 \ifx #1\expandafter \@firstoftwo
 \else \expandafter \@secondoftwo
 \fi
}%
\providecommand \natexlab [1]{#1}%
\providecommand \enquote  [1]{``#1''}%
\providecommand \bibnamefont  [1]{#1}%
\providecommand \bibfnamefont [1]{#1}%
\providecommand \citenamefont [1]{#1}%
\providecommand \href@noop [0]{\@secondoftwo}%
\providecommand \href [0]{\begingroup \@sanitize@url \@href}%
\providecommand \@href[1]{\@@startlink{#1}\@@href}%
\providecommand \@@href[1]{\endgroup#1\@@endlink}%
\providecommand \@sanitize@url [0]{\catcode `\\12\catcode `\$12\catcode
  `\&12\catcode `\#12\catcode `\^12\catcode `\_12\catcode `\%12\relax}%
\providecommand \@@startlink[1]{}%
\providecommand \@@endlink[0]{}%
\providecommand \url  [0]{\begingroup\@sanitize@url \@url }%
\providecommand \@url [1]{\endgroup\@href {#1}{\urlprefix }}%
\providecommand \urlprefix  [0]{URL }%
\providecommand \Eprint [0]{\href }%
\providecommand \doibase [0]{http://dx.doi.org/}%
\providecommand \selectlanguage [0]{\@gobble}%
\providecommand \bibinfo  [0]{\@secondoftwo}%
\providecommand \bibfield  [0]{\@secondoftwo}%
\providecommand \translation [1]{[#1]}%
\providecommand \BibitemOpen [0]{}%
\providecommand \bibitemStop [0]{}%
\providecommand \bibitemNoStop [0]{.\EOS\space}%
\providecommand \EOS [0]{\spacefactor3000\relax}%
\providecommand \BibitemShut  [1]{\csname bibitem#1\endcsname}%
\let\auto@bib@innerbib\@empty
\bibitem [{\citenamefont {Sher}(1989)}]{Sher:1988mj}%
  \BibitemOpen
  \bibfield  {author} {\bibinfo {author} {\bibfnamefont {M.}~\bibnamefont
  {Sher}},\ }\href {\doibase 10.1016/0370-1573(89)90061-6} {\bibfield
  {journal} {\bibinfo  {journal} {Phys. Rept.}\ }\textbf {\bibinfo {volume}
  {179}},\ \bibinfo {pages} {273} (\bibinfo {year} {1989})}\BibitemShut
  {NoStop}%
\bibitem [{\citenamefont {Banks}\ \emph {et~al.}(1973)\citenamefont {Banks},
  \citenamefont {Bender},\ and\ \citenamefont {Wu}}]{BBW}%
  \BibitemOpen
  \bibfield  {author} {\bibinfo {author} {\bibfnamefont {T.}~\bibnamefont
  {Banks}}, \bibinfo {author} {\bibfnamefont {C.~M.}\ \bibnamefont {Bender}}, \
  and\ \bibinfo {author} {\bibfnamefont {T.~T.}\ \bibnamefont {Wu}},\ }\href
  {\doibase 10.1103/PhysRevD.8.3346} {\bibfield  {journal} {\bibinfo  {journal}
  {Phys. Rev. D}\ }\textbf {\bibinfo {volume} {8}},\ \bibinfo {pages} {3346}
  (\bibinfo {year} {1973})}\BibitemShut {NoStop}%
\bibitem [{\citenamefont {Banks}\ and\ \citenamefont
  {Bender}(1973)}]{Banks:1974ij}%
  \BibitemOpen
  \bibfield  {author} {\bibinfo {author} {\bibfnamefont {T.}~\bibnamefont
  {Banks}}\ and\ \bibinfo {author} {\bibfnamefont {C.~M.}\ \bibnamefont
  {Bender}},\ }\href {\doibase 10.1103/PhysRevD.8.3366} {\bibfield  {journal}
  {\bibinfo  {journal} {Phys. Rev.}\ }\textbf {\bibinfo {volume} {D8}},\
  \bibinfo {pages} {3366} (\bibinfo {year} {1973})}\BibitemShut {NoStop}%
\bibitem [{\citenamefont {Kobzarev}\ \emph {et~al.}(1975)\citenamefont
  {Kobzarev}, \citenamefont {Okun},\ and\ \citenamefont
  {Voloshin}}]{Kobzarev:1974cp}%
  \BibitemOpen
  \bibfield  {author} {\bibinfo {author} {\bibfnamefont {I.~{\relax Yu}.}\
  \bibnamefont {Kobzarev}}, \bibinfo {author} {\bibfnamefont {L.~B.}\
  \bibnamefont {Okun}}, \ and\ \bibinfo {author} {\bibfnamefont {M.~B.}\
  \bibnamefont {Voloshin}},\ }\href@noop {} {\bibfield  {journal} {\bibinfo
  {journal} {Sov. J. Nucl. Phys.}\ }\textbf {\bibinfo {volume} {20}},\ \bibinfo
  {pages} {644} (\bibinfo {year} {1975})},\ \bibinfo {note} {[Yad.
  Fiz.20,1229(1974)]}\BibitemShut {NoStop}%
\bibitem [{\citenamefont {Coleman}(1977)}]{Coleman:1977py}%
  \BibitemOpen
  \bibfield  {author} {\bibinfo {author} {\bibfnamefont {S.~R.}\ \bibnamefont
  {Coleman}},\ }\href {\doibase 10.1103/PhysRevD.15.2929,
  10.1103/PhysRevD.16.1248} {\bibfield  {journal} {\bibinfo  {journal} {Phys.
  Rev.}\ }\textbf {\bibinfo {volume} {D15}},\ \bibinfo {pages} {2929} (\bibinfo
  {year} {1977})},\ \bibinfo {note} {[Erratum: Phys.
  Rev.D16,1248(1977)]}\BibitemShut {NoStop}%
\bibitem [{\citenamefont {Callan}\ and\ \citenamefont
  {Coleman}(1977)}]{Callan:1977pt}%
  \BibitemOpen
  \bibfield  {author} {\bibinfo {author} {\bibfnamefont {C.~G.}\ \bibnamefont
  {Callan}, \bibfnamefont {Jr.}}\ and\ \bibinfo {author} {\bibfnamefont
  {S.~R.}\ \bibnamefont {Coleman}},\ }\href {\doibase 10.1103/PhysRevD.16.1762}
  {\bibfield  {journal} {\bibinfo  {journal} {Phys. Rev.}\ }\textbf {\bibinfo
  {volume} {D16}},\ \bibinfo {pages} {1762} (\bibinfo {year}
  {1977})}\BibitemShut {NoStop}%
\bibitem [{\citenamefont {Coleman}\ and\ \citenamefont
  {De~Luccia}(1980)}]{Coleman:1980aw}%
  \BibitemOpen
  \bibfield  {author} {\bibinfo {author} {\bibfnamefont {S.~R.}\ \bibnamefont
  {Coleman}}\ and\ \bibinfo {author} {\bibfnamefont {F.}~\bibnamefont
  {De~Luccia}},\ }\href {\doibase 10.1103/PhysRevD.21.3305} {\bibfield
  {journal} {\bibinfo  {journal} {Phys. Rev.}\ }\textbf {\bibinfo {volume}
  {D21}},\ \bibinfo {pages} {3305} (\bibinfo {year} {1980})}\BibitemShut
  {NoStop}%
\bibitem [{\citenamefont {Hawking}\ and\ \citenamefont
  {Moss}(1982)}]{Hawking:1981fz}%
  \BibitemOpen
  \bibfield  {author} {\bibinfo {author} {\bibfnamefont {S.~W.}\ \bibnamefont
  {Hawking}}\ and\ \bibinfo {author} {\bibfnamefont {I.~G.}\ \bibnamefont
  {Moss}},\ }\href {\doibase 10.1016/0370-2693(82)90946-7} {\bibfield
  {journal} {\bibinfo  {journal} {Phys. Lett.}\ }\textbf {\bibinfo {volume}
  {B110}},\ \bibinfo {pages} {35} (\bibinfo {year} {1982})}\BibitemShut
  {NoStop}%
\bibitem [{\citenamefont {Parke}(1983)}]{Parke:1982pm}%
  \BibitemOpen
  \bibfield  {author} {\bibinfo {author} {\bibfnamefont {S.~J.}\ \bibnamefont
  {Parke}},\ }\href {\doibase 10.1016/0370-2693(83)91376-X} {\bibfield
  {journal} {\bibinfo  {journal} {Phys. Lett.}\ }\textbf {\bibinfo {volume}
  {B121}},\ \bibinfo {pages} {313} (\bibinfo {year} {1983})}\BibitemShut
  {NoStop}%
\bibitem [{\citenamefont {Freivogel}(2011)}]{Freivogel:2011eg}%
  \BibitemOpen
  \bibfield  {author} {\bibinfo {author} {\bibfnamefont {B.}~\bibnamefont
  {Freivogel}},\ }\href {\doibase 10.1088/0264-9381/28/20/204007} {\bibfield
  {journal} {\bibinfo  {journal} {Class. Quant. Grav.}\ }\textbf {\bibinfo
  {volume} {28}},\ \bibinfo {pages} {204007} (\bibinfo {year} {2011})},\
  \Eprint {http://arxiv.org/abs/1105.0244} {arXiv:1105.0244 [hep-th]}
  \BibitemShut {NoStop}%
\bibitem [{\citenamefont {Weinberg}(2007)}]{Weinberg:2006pc}%
  \BibitemOpen
  \bibfield  {author} {\bibinfo {author} {\bibfnamefont {E.~J.}\ \bibnamefont
  {Weinberg}},\ }\href {\doibase 10.1103/PhysRevLett.98.251303} {\bibfield
  {journal} {\bibinfo  {journal} {Phys. Rev. Lett.}\ }\textbf {\bibinfo
  {volume} {98}},\ \bibinfo {pages} {251303} (\bibinfo {year} {2007})},\
  \Eprint {http://arxiv.org/abs/hep-th/0612146} {arXiv:hep-th/0612146 [hep-th]}
  \BibitemShut {NoStop}%
\bibitem [{\citenamefont {Masoumi}\ and\ \citenamefont
  {Weinberg}(2012)}]{Masoumi:2012yy}%
  \BibitemOpen
  \bibfield  {author} {\bibinfo {author} {\bibfnamefont {A.}~\bibnamefont
  {Masoumi}}\ and\ \bibinfo {author} {\bibfnamefont {E.~J.}\ \bibnamefont
  {Weinberg}},\ }\href {\doibase 10.1103/PhysRevD.86.104029} {\bibfield
  {journal} {\bibinfo  {journal} {Phys. Rev.}\ }\textbf {\bibinfo {volume}
  {D86}},\ \bibinfo {pages} {104029} (\bibinfo {year} {2012})},\ \Eprint
  {http://arxiv.org/abs/1207.3717} {arXiv:1207.3717 [hep-th]} \BibitemShut
  {NoStop}%
\bibitem [{\citenamefont {Garriga}\ and\ \citenamefont
  {Megevand}(2004)}]{Garriga:2004nm}%
  \BibitemOpen
  \bibfield  {author} {\bibinfo {author} {\bibfnamefont {J.}~\bibnamefont
  {Garriga}}\ and\ \bibinfo {author} {\bibfnamefont {A.}~\bibnamefont
  {Megevand}},\ }\bibfield  {booktitle} {\emph {\bibinfo {booktitle} {{The
  early universe: Confronting theory with observations. Proceedings, 8th
  Workshop, Peyresq Physics 8, Peyresq, France, June 21-27, 2003}}},\ }\href
  {\doibase 10.1023/B:IJTP.0000048178.69097.fb} {\bibfield  {journal} {\bibinfo
   {journal} {Int. J. Theor. Phys.}\ }\textbf {\bibinfo {volume} {43}},\
  \bibinfo {pages} {883} (\bibinfo {year} {2004})},\ \Eprint
  {http://arxiv.org/abs/hep-th/0404097} {arXiv:hep-th/0404097 [hep-th]}
  \BibitemShut {NoStop}%
\bibitem [{\citenamefont {Masoumi}\ \emph {et~al.}(2017)\citenamefont
  {Masoumi}, \citenamefont {Olum},\ and\ \citenamefont
  {Shlaer}}]{Masoumi:2016wot}%
  \BibitemOpen
  \bibfield  {author} {\bibinfo {author} {\bibfnamefont {A.}~\bibnamefont
  {Masoumi}}, \bibinfo {author} {\bibfnamefont {K.~D.}\ \bibnamefont {Olum}}, \
  and\ \bibinfo {author} {\bibfnamefont {B.}~\bibnamefont {Shlaer}},\ }\href
  {\doibase 10.1088/1475-7516/2017/01/051} {\bibfield  {journal} {\bibinfo
  {journal} {JCAP}\ }\textbf {\bibinfo {volume} {1701}},\ \bibinfo {pages}
  {051} (\bibinfo {year} {2017})},\ \Eprint {http://arxiv.org/abs/1610.06594}
  {arXiv:1610.06594 [gr-qc]} \BibitemShut {NoStop}%
\bibitem [{\citenamefont {Clifton}\ \emph {et~al.}(2012)\citenamefont
  {Clifton}, \citenamefont {Ferreira}, \citenamefont {Padilla},\ and\
  \citenamefont {Skordis}}]{Clifton:2011jh}%
  \BibitemOpen
  \bibfield  {author} {\bibinfo {author} {\bibfnamefont {T.}~\bibnamefont
  {Clifton}}, \bibinfo {author} {\bibfnamefont {P.~G.}\ \bibnamefont
  {Ferreira}}, \bibinfo {author} {\bibfnamefont {A.}~\bibnamefont {Padilla}}, \
  and\ \bibinfo {author} {\bibfnamefont {C.}~\bibnamefont {Skordis}},\ }\href
  {\doibase 10.1016/j.physrep.2012.01.001} {\bibfield  {journal} {\bibinfo
  {journal} {Phys. Rept.}\ }\textbf {\bibinfo {volume} {513}},\ \bibinfo
  {pages} {1} (\bibinfo {year} {2012})},\ \Eprint
  {http://arxiv.org/abs/1106.2476} {arXiv:1106.2476 [astro-ph.CO]} \BibitemShut
  {NoStop}%
\bibitem [{\citenamefont {Joyce}\ \emph {et~al.}(2015)\citenamefont {Joyce},
  \citenamefont {Jain}, \citenamefont {Khoury},\ and\ \citenamefont
  {Trodden}}]{Joyce:2014kja}%
  \BibitemOpen
  \bibfield  {author} {\bibinfo {author} {\bibfnamefont {A.}~\bibnamefont
  {Joyce}}, \bibinfo {author} {\bibfnamefont {B.}~\bibnamefont {Jain}},
  \bibinfo {author} {\bibfnamefont {J.}~\bibnamefont {Khoury}}, \ and\ \bibinfo
  {author} {\bibfnamefont {M.}~\bibnamefont {Trodden}},\ }\href {\doibase
  10.1016/j.physrep.2014.12.002} {\bibfield  {journal} {\bibinfo  {journal}
  {Phys. Rept.}\ }\textbf {\bibinfo {volume} {568}},\ \bibinfo {pages} {1}
  (\bibinfo {year} {2015})},\ \Eprint {http://arxiv.org/abs/1407.0059}
  {arXiv:1407.0059 [astro-ph.CO]} \BibitemShut {NoStop}%
\bibitem [{\citenamefont {Brown}\ \emph {et~al.}(2007)\citenamefont {Brown},
  \citenamefont {Sarangi}, \citenamefont {Shlaer},\ and\ \citenamefont
  {Weltman}}]{Brown:2007ce}%
  \BibitemOpen
  \bibfield  {author} {\bibinfo {author} {\bibfnamefont {A.~R.}\ \bibnamefont
  {Brown}}, \bibinfo {author} {\bibfnamefont {S.}~\bibnamefont {Sarangi}},
  \bibinfo {author} {\bibfnamefont {B.}~\bibnamefont {Shlaer}}, \ and\ \bibinfo
  {author} {\bibfnamefont {A.}~\bibnamefont {Weltman}},\ }\href {\doibase
  10.1103/PhysRevLett.99.161601} {\bibfield  {journal} {\bibinfo  {journal}
  {Phys. Rev. Lett.}\ }\textbf {\bibinfo {volume} {99}},\ \bibinfo {pages}
  {161601} (\bibinfo {year} {2007})},\ \Eprint {http://arxiv.org/abs/0706.0485}
  {arXiv:0706.0485 [hep-th]} \BibitemShut {NoStop}%
\bibitem [{\citenamefont {Cohen}(1966)}]{Cohen:1966}%
  \BibitemOpen
  \bibfield  {author} {\bibinfo {author} {\bibfnamefont {L.}~\bibnamefont
  {Cohen}},\ }\href {\doibase 10.1063/1.1931206} {\bibfield  {journal}
  {\bibinfo  {journal} {Journal of Mathematical Physics}\ }\textbf {\bibinfo
  {volume} {7}},\ \bibinfo {pages} {781} (\bibinfo {year} {1966})}\BibitemShut
  {NoStop}%
\bibitem [{\citenamefont {Mizrahi}(1981)}]{Miz81}%
  \BibitemOpen
  \bibfield  {author} {\bibinfo {author} {\bibfnamefont {M.~M.}\ \bibnamefont
  {Mizrahi}},\ }\href {\doibase 10.1063/1.524739} {\bibfield  {journal}
  {\bibinfo  {journal} {Journal of Mathematical Physics}\ }\textbf {\bibinfo
  {volume} {22}},\ \bibinfo {pages} {102} (\bibinfo {year} {1981})}\BibitemShut
  {NoStop}%
\bibitem [{\citenamefont {Mizrahi}(1977)}]{Miz77}%
  \BibitemOpen
  \bibfield  {author} {\bibinfo {author} {\bibfnamefont {M.~M.}\ \bibnamefont
  {Mizrahi}},\ }\href {\doibase 10.1063/1.523308} {\bibfield  {journal}
  {\bibinfo  {journal} {Journal of Mathematical Physics}\ }\textbf {\bibinfo
  {volume} {18}},\ \bibinfo {pages} {786} (\bibinfo {year} {1977})}\BibitemShut
  {NoStop}%
\bibitem [{\citenamefont {Brown}(1972)}]{Brown72}%
  \BibitemOpen
  \bibfield  {author} {\bibinfo {author} {\bibfnamefont {L.~S.}\ \bibnamefont
  {Brown}},\ }\href {\doibase 10.1119/1.1986554} {\bibfield  {journal}
  {\bibinfo  {journal} {American Journal of Physics}\ }\textbf {\bibinfo
  {volume} {40}},\ \bibinfo {pages} {371} (\bibinfo {year} {1972})}\BibitemShut
  {NoStop}%
\bibitem [{\citenamefont {Bitar}\ and\ \citenamefont
  {Chang}(1978)}]{Bitar:1978vx}%
  \BibitemOpen
  \bibfield  {author} {\bibinfo {author} {\bibfnamefont {K.~M.}\ \bibnamefont
  {Bitar}}\ and\ \bibinfo {author} {\bibfnamefont {S.-J.}\ \bibnamefont
  {Chang}},\ }\href {\doibase 10.1103/PhysRevD.18.435} {\bibfield  {journal}
  {\bibinfo  {journal} {Phys. Rev.}\ }\textbf {\bibinfo {volume} {D18}},\
  \bibinfo {pages} {435} (\bibinfo {year} {1978})}\BibitemShut {NoStop}%
\bibitem [{\citenamefont {Copeland}\ \emph {et~al.}(2008)\citenamefont
  {Copeland}, \citenamefont {Padilla},\ and\ \citenamefont
  {Saffin}}]{Copeland:2007qf}%
  \BibitemOpen
  \bibfield  {author} {\bibinfo {author} {\bibfnamefont {E.~J.}\ \bibnamefont
  {Copeland}}, \bibinfo {author} {\bibfnamefont {A.}~\bibnamefont {Padilla}}, \
  and\ \bibinfo {author} {\bibfnamefont {P.~M.}\ \bibnamefont {Saffin}},\
  }\href {\doibase 10.1088/1126-6708/2008/01/066} {\bibfield  {journal}
  {\bibinfo  {journal} {JHEP}\ }\textbf {\bibinfo {volume} {01}},\ \bibinfo
  {pages} {066} (\bibinfo {year} {2008})},\ \Eprint
  {http://arxiv.org/abs/0709.0261} {arXiv:0709.0261 [hep-th]} \BibitemShut
  {NoStop}%
\bibitem [{\citenamefont {Masoumi}(2013)}]{Masoumi:2015psa}%
  \BibitemOpen
  \bibfield  {author} {\bibinfo {author} {\bibfnamefont {A.}~\bibnamefont
  {Masoumi}},\ }\emph {\bibinfo {title} {{Topics in vacuum decay}}},\ \href
  {https://inspirehep.net/record/1372763/files/arXiv:1505.06397.pdf} {Ph.D.
  thesis},\ \bibinfo  {school} {Columbia U} (\bibinfo {year} {2013}),\ \Eprint
  {http://arxiv.org/abs/1505.06397} {arXiv:1505.06397 [hep-th]} \BibitemShut
  {NoStop}%
\bibitem [{\citenamefont {Andreassen}\ \emph {et~al.}(2016)\citenamefont
  {Andreassen}, \citenamefont {Farhi}, \citenamefont {Frost},\ and\
  \citenamefont {Schwartz}}]{Andreassen:2016cvx}%
  \BibitemOpen
  \bibfield  {author} {\bibinfo {author} {\bibfnamefont {A.}~\bibnamefont
  {Andreassen}}, \bibinfo {author} {\bibfnamefont {D.}~\bibnamefont {Farhi}},
  \bibinfo {author} {\bibfnamefont {W.}~\bibnamefont {Frost}}, \ and\ \bibinfo
  {author} {\bibfnamefont {M.~D.}\ \bibnamefont {Schwartz}},\ }\href@noop {} {\
   (\bibinfo {year} {2016})},\ \Eprint {http://arxiv.org/abs/1604.06090}
  {arXiv:1604.06090 [hep-th]} \BibitemShut {NoStop}%
\bibitem [{\citenamefont {Coleman}\ \emph {et~al.}(1978)\citenamefont
  {Coleman}, \citenamefont {Glaser},\ and\ \citenamefont
  {Martin}}]{Coleman:1977th}%
  \BibitemOpen
  \bibfield  {author} {\bibinfo {author} {\bibfnamefont {S.~R.}\ \bibnamefont
  {Coleman}}, \bibinfo {author} {\bibfnamefont {V.}~\bibnamefont {Glaser}}, \
  and\ \bibinfo {author} {\bibfnamefont {A.}~\bibnamefont {Martin}},\ }\href
  {\doibase 10.1007/BF01609421} {\bibfield  {journal} {\bibinfo  {journal}
  {Commun. Math. Phys.}\ }\textbf {\bibinfo {volume} {58}},\ \bibinfo {pages}
  {211} (\bibinfo {year} {1978})}\BibitemShut {NoStop}%
\bibitem [{\citenamefont {Nicolis}\ \emph {et~al.}(2009)\citenamefont
  {Nicolis}, \citenamefont {Rattazzi},\ and\ \citenamefont
  {Trincherini}}]{Nicolis:2008in}%
  \BibitemOpen
  \bibfield  {author} {\bibinfo {author} {\bibfnamefont {A.}~\bibnamefont
  {Nicolis}}, \bibinfo {author} {\bibfnamefont {R.}~\bibnamefont {Rattazzi}}, \
  and\ \bibinfo {author} {\bibfnamefont {E.}~\bibnamefont {Trincherini}},\
  }\href {\doibase 10.1103/PhysRevD.79.064036} {\bibfield  {journal} {\bibinfo
  {journal} {Phys. Rev.}\ }\textbf {\bibinfo {volume} {D79}},\ \bibinfo {pages}
  {064036} (\bibinfo {year} {2009})},\ \Eprint {http://arxiv.org/abs/0811.2197}
  {arXiv:0811.2197 [hep-th]} \BibitemShut {NoStop}%
\bibitem [{\citenamefont {Horndeski}(1974)}]{Horndeski:1974wa}%
  \BibitemOpen
  \bibfield  {author} {\bibinfo {author} {\bibfnamefont {G.~W.}\ \bibnamefont
  {Horndeski}},\ }\href {\doibase 10.1007/BF01807638} {\bibfield  {journal}
  {\bibinfo  {journal} {Int. J. Theor. Phys.}\ }\textbf {\bibinfo {volume}
  {10}},\ \bibinfo {pages} {363} (\bibinfo {year} {1974})}\BibitemShut
  {NoStop}%
\bibitem [{\citenamefont {Deffayet}\ \emph {et~al.}(2011)\citenamefont
  {Deffayet}, \citenamefont {Gao}, \citenamefont {Steer},\ and\ \citenamefont
  {Zahariade}}]{Deffayet:2011gz}%
  \BibitemOpen
  \bibfield  {author} {\bibinfo {author} {\bibfnamefont {C.}~\bibnamefont
  {Deffayet}}, \bibinfo {author} {\bibfnamefont {X.}~\bibnamefont {Gao}},
  \bibinfo {author} {\bibfnamefont {D.~A.}\ \bibnamefont {Steer}}, \ and\
  \bibinfo {author} {\bibfnamefont {G.}~\bibnamefont {Zahariade}},\ }\href
  {\doibase 10.1103/PhysRevD.84.064039} {\bibfield  {journal} {\bibinfo
  {journal} {Phys. Rev.}\ }\textbf {\bibinfo {volume} {D84}},\ \bibinfo {pages}
  {064039} (\bibinfo {year} {2011})},\ \Eprint {http://arxiv.org/abs/1103.3260}
  {arXiv:1103.3260 [hep-th]} \BibitemShut {NoStop}%
\bibitem [{\citenamefont {Ostrogradsky}(1850)}]{Ostrogradsky:1850fid}%
  \BibitemOpen
  \bibfield  {author} {\bibinfo {author} {\bibfnamefont {M.}~\bibnamefont
  {Ostrogradsky}},\ }\href@noop {} {\bibfield  {journal} {\bibinfo  {journal}
  {Mem. Acad. St. Petersbourg}\ }\textbf {\bibinfo {volume} {6}},\ \bibinfo
  {pages} {385} (\bibinfo {year} {1850})}\BibitemShut {NoStop}%
\bibitem [{\citenamefont {Woodard}(2007)}]{Woodard:2006nt}%
  \BibitemOpen
  \bibfield  {author} {\bibinfo {author} {\bibfnamefont {R.~P.}\ \bibnamefont
  {Woodard}},\ }\href {\doibase 10.1007/978-3-540-71013-4_14} {\bibfield
  {journal} {\bibinfo  {journal} {Lect.Notes Phys.}\ }\textbf {\bibinfo
  {volume} {720}},\ \bibinfo {pages} {403} (\bibinfo {year} {2007})},\ \Eprint
  {http://arxiv.org/abs/astro-ph/0601672} {arXiv:astro-ph/0601672 [astro-ph]}
  \BibitemShut {NoStop}%
\bibitem [{\citenamefont {de~Rham}\ and\ \citenamefont
  {Matas}(2016)}]{deRham:2016wji}%
  \BibitemOpen
  \bibfield  {author} {\bibinfo {author} {\bibfnamefont {C.}~\bibnamefont
  {de~Rham}}\ and\ \bibinfo {author} {\bibfnamefont {A.}~\bibnamefont
  {Matas}},\ }\href {\doibase 10.1088/1475-7516/2016/06/041} {\bibfield
  {journal} {\bibinfo  {journal} {JCAP}\ }\textbf {\bibinfo {volume} {1606}},\
  \bibinfo {pages} {041} (\bibinfo {year} {2016})},\ \Eprint
  {http://arxiv.org/abs/1604.08638} {arXiv:1604.08638 [hep-th]} \BibitemShut
  {NoStop}%
\bibitem [{\citenamefont {Gleyzes}\ \emph {et~al.}(2015)\citenamefont
  {Gleyzes}, \citenamefont {Langlois}, \citenamefont {Piazza},\ and\
  \citenamefont {Vernizzi}}]{Gleyzes:2014dya}%
  \BibitemOpen
  \bibfield  {author} {\bibinfo {author} {\bibfnamefont {J.}~\bibnamefont
  {Gleyzes}}, \bibinfo {author} {\bibfnamefont {D.}~\bibnamefont {Langlois}},
  \bibinfo {author} {\bibfnamefont {F.}~\bibnamefont {Piazza}}, \ and\ \bibinfo
  {author} {\bibfnamefont {F.}~\bibnamefont {Vernizzi}},\ }\href {\doibase
  10.1103/PhysRevLett.114.211101} {\bibfield  {journal} {\bibinfo  {journal}
  {Phys. Rev. Lett.}\ }\textbf {\bibinfo {volume} {114}},\ \bibinfo {pages}
  {211101} (\bibinfo {year} {2015})},\ \Eprint {http://arxiv.org/abs/1404.6495}
  {arXiv:1404.6495 [hep-th]} \BibitemShut {NoStop}%
\bibitem [{\citenamefont {Zumalacárregui}\ and\ \citenamefont
  {García-Bellido}(2014)}]{Zumalacarregui:2013pma}%
  \BibitemOpen
  \bibfield  {author} {\bibinfo {author} {\bibfnamefont {M.}~\bibnamefont
  {Zumalacárregui}}\ and\ \bibinfo {author} {\bibfnamefont {J.}~\bibnamefont
  {García-Bellido}},\ }\href {\doibase 10.1103/PhysRevD.89.064046} {\bibfield
  {journal} {\bibinfo  {journal} {Phys. Rev.}\ }\textbf {\bibinfo {volume}
  {D89}},\ \bibinfo {pages} {064046} (\bibinfo {year} {2014})},\ \Eprint
  {http://arxiv.org/abs/1308.4685} {arXiv:1308.4685 [gr-qc]} \BibitemShut
  {NoStop}%
\bibitem [{\citenamefont {Crisostomi}\ \emph {et~al.}(2016)\citenamefont
  {Crisostomi}, \citenamefont {Koyama},\ and\ \citenamefont
  {Tasinato}}]{Crisostomi:2016czh}%
  \BibitemOpen
  \bibfield  {author} {\bibinfo {author} {\bibfnamefont {M.}~\bibnamefont
  {Crisostomi}}, \bibinfo {author} {\bibfnamefont {K.}~\bibnamefont {Koyama}},
  \ and\ \bibinfo {author} {\bibfnamefont {G.}~\bibnamefont {Tasinato}},\
  }\href {\doibase 10.1088/1475-7516/2016/04/044} {\bibfield  {journal}
  {\bibinfo  {journal} {JCAP}\ }\textbf {\bibinfo {volume} {1604}},\ \bibinfo
  {pages} {044} (\bibinfo {year} {2016})},\ \Eprint
  {http://arxiv.org/abs/1602.03119} {arXiv:1602.03119 [hep-th]} \BibitemShut
  {NoStop}%
\bibitem [{\citenamefont {Ben~Achour}\ \emph {et~al.}(2016)\citenamefont
  {Ben~Achour}, \citenamefont {Langlois},\ and\ \citenamefont
  {Noui}}]{Achour:2016rkg}%
  \BibitemOpen
  \bibfield  {author} {\bibinfo {author} {\bibfnamefont {J.}~\bibnamefont
  {Ben~Achour}}, \bibinfo {author} {\bibfnamefont {D.}~\bibnamefont
  {Langlois}}, \ and\ \bibinfo {author} {\bibfnamefont {K.}~\bibnamefont
  {Noui}},\ }\href {\doibase 10.1103/PhysRevD.93.124005} {\bibfield  {journal}
  {\bibinfo  {journal} {Phys. Rev.}\ }\textbf {\bibinfo {volume} {D93}},\
  \bibinfo {pages} {124005} (\bibinfo {year} {2016})},\ \Eprint
  {http://arxiv.org/abs/1602.08398} {arXiv:1602.08398 [gr-qc]} \BibitemShut
  {NoStop}%
\bibitem [{\citenamefont {Sivanesan}(2012)}]{Sivanesan:2011kw}%
  \BibitemOpen
  \bibfield  {author} {\bibinfo {author} {\bibfnamefont {V.}~\bibnamefont
  {Sivanesan}},\ }\href {\doibase 10.1103/PhysRevD.85.084018} {\bibfield
  {journal} {\bibinfo  {journal} {Phys. Rev.}\ }\textbf {\bibinfo {volume}
  {D85}},\ \bibinfo {pages} {084018} (\bibinfo {year} {2012})},\ \Eprint
  {http://arxiv.org/abs/1111.3558} {arXiv:1111.3558 [hep-th]} \BibitemShut
  {NoStop}%
\end{thebibliography}%

\end{document}